# Data-driven competitive facilitative tree interactions and their implications on nature-based solutions

Aristides Moustakas[1,*], Ioannis N. Daliakopoulos[2] and Tim. G. Benton[3]

1. Institute for Applied Data Analytics, Universiti Brunei Darussalam, Jalan Tungku Link, Gadong BE 1410, Brunei

2. Department of Agriculture, Technological Educational Institute of Crete, Heraklion, Greece

3. School of Biology, University of Leeds, Leeds LS2 9JT, UK

- Corresponding author:
  Aris Moustakas

  Email: arismoustakas@gmail.com

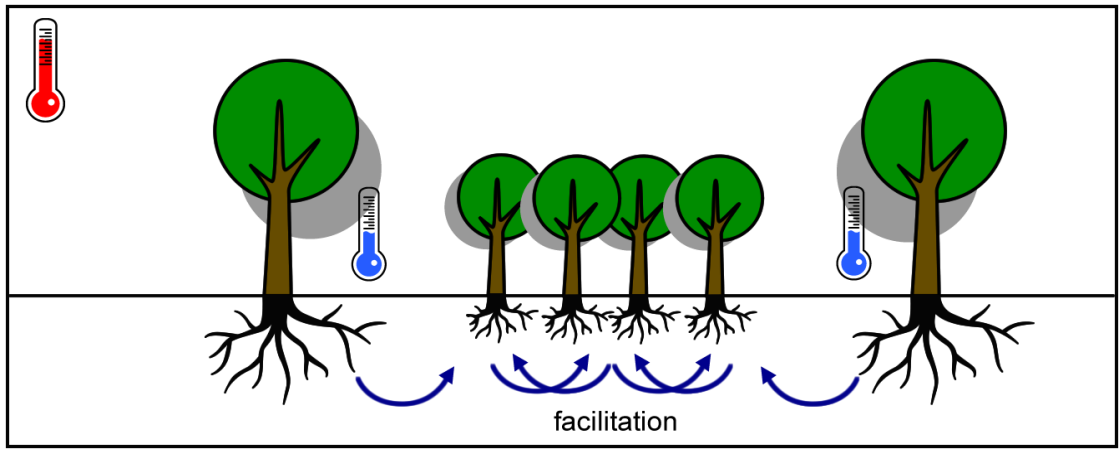

**Highlights**

(i) Data-driven approaches may provide new insights into understanding facilitation

(ii) Facilitation & competition coexist depending on the spatial scale.

(iii) Within a spatial scale there are competitive years and facilitative years.

(iv) Tree size is more important predictor than density in tree survival

(v) Large trees moderate surface temperature more than vegetation density

**Abstract**

Spatio-temporal data are more ubiquitous and richer than even before and the availability of such data poses great challenges in data analytics. Ecological facilitation, the positive effect of density of individuals on the individual's survival across a stress gradient, is a complex phenomenon. A large number of tree individuals coupled with soil moisture, temperature, and water stress data across a long temporal period were followed. Data-driven analysis in the absence of hypothesis was performed. Information theoretic analysis of multiple statistical models was employed in order to quantify the best data-driven index of vegetation density and spatial scale of interactions. Sequentially, tree survival was quantified as a function of the size of the individual, vegetation density, and time at the optimal spatial interaction scale. Land surface temperature and soil moisture were also statistically explained by tree size, density, and time. Results indicated that in space both facilitation and competition co-exist in the same ecosystem and the sign and magnitude of this depend on the spatial scale. Overall, within the optimal data-driven spatial scale, tree survival was best explained by the interaction between density and year, sifting overall from facilitation to competition through time. However, small sized trees were always facilitated by increased densities, while large sized trees had either negative or no density effects. Tree size was more important predictor than density in survival and this has implications for nature-based solutions: maintaining large tree individuals or planting species that can become large-sized can safeguard against tree-less areas by promoting survival at long time periods through harsh environmental conditions. Large trees had also a significant effect in moderating land surface temperature and this effect was higher than the one of vegetation density on temperature.

**Keywords**

Environmental informatics; ecological facilitation; eco-hydrology; density effects; soil moisture; land surface temperature; savannas

**Introduction**

With the rapid development of smart sensors, social networks, as well as digital maps and remotely-sensed imagery, spatio-temporal data are more ubiquitous and richer than ever before (Fayyad et al., 1996; Gupta et al., 1997; Moustakas, 2017). The volume of such (big) data creates great challenges in the handling, visualizing, and analysing (Chen and Zhang, 2014; Jagadish et al., 2014; Moustakas and Katsanevakis, 2018). These challenges have generated the necessity of new interdisciplinary fields between statistics, computer science, and the field of the data domain, potentially providing a paradigm shift in science (Kitchin, 2014), with data-driven approaches (Deluigi et al., 2017; Hong et al., 2018; Levi et al., 2018; Yin et al., 2018).

Since its advent, remote sensing has provided important coverage, mapping and classification of land-cover features, such as vegetation, soil, and water (Lillesand et al., 2014). Remote sensing is giving us unprecedented access to data between ecosystems and climate (Kerr and Ostrovsky, 2003; Reichstein et al., 2007), allowing us to explore ecological effects which are weak at the individual scale but important in determining ecosystem-level properties. Within the field of ecology, the availability of remotely sensed imagery has huge potential for addressing ecological questions at scales unimaginable in the past (Xu et al., 2015). One area is the ability to look at patterns of survival between plants over large scales and multiple time steps (Moustakas et al., 2010). This is important because the way trees interact and survive determines a range of ecosystem services and thus has implications for nature-based solutions (Baró and Gómez-Baggethun, 2017).

Often under-recognised in the past, positive (facilitation) and negative (competition) density plant interactions are now considered to have serious implications for population dynamics and ecosystem function (Brooker et al., 2008). There are several definitions of facilitation (Wright et al., 2017) as well as interactions occurring between different types of plant, such as tree-tree, shrub-tree, tree-grass or woody species-grass, grass-seedling, and seedling-adult trees. Here, our focus is on tree survival, so we use the term 'facilitation' as increased chance of woody species survival (tree-tree and shrub-tree, i.e. among woody species only, thereafter tree) with increasing number of individual neighbours or canopy cover within a defined neighbourhood. Facilitative-competitive interactions have often been investigated in terms of the stress gradient hypothesis (Bertness and Callaway, 1994), which predicts an increase of positive interactions (facilitation) with increasing environmental stress (Bertness and Callaway, 1994; Blaser et al., 2013; Dohn et al., 2013). Apart from the interest that facilitation exhibits from the perspective of basic biological knowledge, it has serious implications for soil surface water, degraded land restoration (Antonio et al., 2018; Víctor et al., 2017), as well as for agriculture and food security (Li et al., 2014; Moschitz et al., 2015). To that end, understanding the interplay between trees and the way they ameliorate their biotic and physical environment (Yu and D'Odorico, 2017) as well as climate and soil, is critical for ecosystem management, agricultural planning, and water management (Davis et al., 2017).

Negative density effects on plant life-histories have been reported to switch to positive effects along a stress gradient, with precipitation as the most commonly reported stress (Noemí et al., 2016). However, there are several other key stressors that can include, among others, elevation (Cavieres et al., 2006; Choler et al., 2001), grazing (Smit et al., 2007), fire (Moustakas, 2015), and temperature (Callaway and King, 1996). In addition, the strength and sign of density effects depends on both on the spatial and temporal scales. Regarding spatial scales, there are cases where competition and facilitation coexist in the same ecosystem (Staver, 2018), with finer scale facilitation and coarser scale competition (Riginos et al., 2009), or the inverse (van de Koppel et al., 2006).

In terms of temporal scales it has been documented that depending on the daily environmental conditions the same individual plants can compete or facilitate depending on water availability and temperature (Wright et al., 2015). These relationships can change over time, in some years facilitation or competition may dominate. Even in years, where, on average, facilitation may dominate, there may be days that competition prevails. Thus, this provides a challenge in selecting the temporal scale for analysis. Adding to the complexity of the problem, positive or negative density effects depend also on the density *per se* as a stress gradient with studies reporting that facilitation peaking at intermediate densities (Dickie et al., 2005) and other studies reporting that higher densities would increase both competition and facilitation determined by the environmental stress gradient (Wright et al., 2015). Part of the complexity (Veblen, 2008; Wright et al., 2013) and lack of a clear picture may derive from the fact that the definition of the spatial scale (neighbourhood) of local interactions becomes a crucial determinant of the power to detect effects at different scales (Bradter et al., 2013; Gunton and Pöyry, 2016); (Fig. 1).

In this study we employ hypothesis-free big data analytics of the impact on tree survival by other tree individuals across 61 years. We integrate a previously published tree dataset with a large number of tree individuals across a long temporal replicate (Moustakas et al., 2006; Moustakas et al., 2008) with temperature, soil water, and water stress data. All data are derived by remote sensing. Doing so we retrospectively integrate tree data with hydrological and climatic data that were not previously available, as the first two time replicates of the tree data were derived in periods when satellites were not available (aerial photos were used instead), and thus the hydrological and temperature data could not be extracted. Past studies in facilitation have mainly been hypothesis-driven. The problem is that when we make a hypothesis, we become attached to it (Chamberlin, 1897; Platt, 1964). We explicitly refrained from formulating any hypotheses; instead we performed data-driven analysis, making the implicit hypothesis that an underlying dependence between collected data can be objectively mined (van Helden, 2013). Data-driven approaches are not competitive to hypothesis-led studies in scientific knowledge discovery but are complementary and iterative with them (Kell and Oliver, 2004). To that end, we initially perform data-driven selection of the best index of tree density as well as the spatial scale of interactions. We then study survival as a function of density, size of the individual, and year as well as their interactions. We sought to quantify the data-driven index of density as well as scale of spatial interactions. We sequentially investigated the spatio-temporal patterns of density effects.

**Methods**

*Study area*

Plots (N=7) are located in semi-arid savanna in the Southern Kalahari near the city of Kimberley, South Africa, covering a total area of ~700 ha. A satellite view of the plots is provided in Fig. 1a, and a ground view in Fig. 1b. Plots extend between 28°55"00' S, 24°77"40' E and 28°65"00' S, 24°88"90' E. Rain is seasonal, falling mainly between December – February (summer months) (Moustakas et al., 2006). Mean annual precipitation is 411 mm (St.Dev = 132), while summer mean maximum daily temperature is 32 $^oC$, and winter mean minimum daily temperature is 3 $^oC$ (Moustakas et al., 2006). The soil consists of mainly Hutton (haplic arenosol) type soil and its depth exceeds 2 m (Moustakas et al., 2006). The main tree species present in the plots are *Acacia erioloba*, *Acacia hebeclada*, *Acacia tortilis*, *Grewia flava*, and *Tarchonanthus camphorate,* with *A. erioloba* being by far the most dominant species, as precipitation is scarce and the sandy soil in the area allows deep rooted species to access permanent deep-soil aquifers (Moustakas et al., 2006). In general the study plots had grazing and some browsing, whereas anthropogenic disturbances and land-uses were minimal (Moustakas et al., 2006).

*Normalized Difference Vegetation Index (NDVI)*

The Normalized Difference Vegetation Index (NDVI) ((Deering and Haas, 1980; Tucker, 1979) is commonly used to represent the level or intensity of vegetation activity. It is based on a simple ratio between the near infrared (*NIR*) and red (*R*) spectral bands, which characterize leaves development and photosynthesis, respectively (Daliakopoulos et al., 2009). The MODIS satellite data with a ground resolution of 250 m was used over the plots.

$$NDVI = \frac{\rho_{NIR} - \rho_R}{\rho_{NIR} + \rho_R} \qquad (1)$$

where $\rho_i$ is the reflectance for the red and near infrared bands, denoted by subscripts *R* and *NIR*, respectively.

*Land Surface Temperature (LST)*

Land surface temperature (LST) is a significant parameter in exploring the exchange of surface matter, surface energy balance and surface physical and chemical processes and is currently widely used in soil, hydrology, biology and geochemistry (Deng et al., 2018; Hao et al., 2016; Tomlinson et al., 2011). Landsat 4 and its successors have one or two thermal bands, and they offer the possibility of obtaining LST estimates at 30 m resolution. Emissivity data must be estimated or alternatively obtained from secondary sources (Muro et al., 2016). A common way of estimating emissivity is the NDVI threshold method, which is based on the statistical relationship existing between thermal and visible and near-infrared bands ((Deng et al., 2018; Sobrino et al., 2004)Deng et al., 2018; Sobrino et al., 2004). The relationship has also been reported to yield higher accuracies in arid areas (Sobrino et al., 2008), therefore its use in the case study is ideal. Here we estimate LST from a single thermal channel using the generalized SC algorithm proposed by (Jiménez-Muñoz et al., 2009; Jiménez-Muñoz and Sobrino, 2003) that is applicable to the TIR channel of Landsat 5 relying on the estimation of the so-called atmospheric functions (AFs), which were assumed to be dependent only on atmospheric water vapor content and land surface emissivity.

*Soil Moisture Index (SMI)*

Soil moisture retained a great deal of attention during recent years, with several relevant indicators being proposed using a wide range of completely different methods and sensors (Kerenyi and Putsay, 2000; Vlassova et al., 2014). Given the NDVI and the Land Surface Temperature Ts [°K] of each pixel, one can define the SMI index for a large enough land dataset (i.e. an entire satellite scene). Ts/NVDI values can be used to derive

$$Ts_{max} = a_d NDVI + b_d \qquad (2)$$
$$Ts_{min} = a_w NDVI + b_w \qquad (3)$$
$$SMI = \frac{Ts_{max} - Ts}{Ts_{max} - Ts_{min}} \qquad (4)$$

where Ts is the Land Surface Temperature [°K] of each pixel, and *a* [°K] and *b* [°K] are the slope and intercept values of the dry and wet edge as denoted by subscripts *d* and *w*. On a conceptual level, the scatter plots of the Ts/NDVI space can be enveloped in a triangular (Carlson et al., 1994) or a trapezium shape (Moran et al., 1994) where the upper sloping edge is defined as the dry edge ($Ts_{max}$), and the lower sloping edge is defined as the wet edge ($Ts_{min}$), since they represent extreme conditions of soil moisture and evapotranspiration.

*Tree data*

A previously published (Moustakas et al., 2006; Moustakas et al., 2008) long-term (1940 to 2001 over 5 time snapshots) tree data set covering over 20,000 individuals within the study area plots was used. The study species are evergreen and thus their NDVI reflectance has very low inter-annual variation (Moustakas et al., 2008). Hence, the fact that the aerial photos have not been taken during the same month each year does not introduce significant bias in the projected tree size (Moustakas et al., 2008). There was negligible browsing, anthropogenic disturbances, or tree diseases in the study plots (Moustakas et al., 2006). As a result, the tree canopy size (as estimated from remote sensing) is not biased by these causes (Moustakas et al., 2006; Moustakas et al., 2008).

For the identification and multi-temporal analysis of the trees, black-and-white aerial photographs of the area taken in 1940, 1964, 1984, and 1993, and an IKONOS satellite image taken in 2001 were used. Every individual tree was identified and followed from 1940 to the next available photo till 2001. The spatial resolution of the aerial photos was 2 m, whereas that of the IKONOS satellite image was 1 m. Since both satellite and aerial photos are used in the tree database, the spatial resolution was set to 2 m; as a result, trees with canopy diameter of minimum 2 m are clustered. Thus, assuming a cyclical projected canopy, the minimum projected canopy area (tree size) recorded is approximately:

$$\pi R^2 = \pi \left(\frac{2 \text{ m}}{2}\right)^2 = 3.14 \text{ m}^2 \qquad (6)$$

In order to avoid the tree delineation error, we excluded from the analysis all trees appearing to have canopy area larger than 350 m², which was the maximum recorded and verified during fieldwork (Moustakas et al., 2006). The study area analyzed in this paper contained 16,331 tree individuals in total across years. Field work for comparing patterns in the classification and easily visible ground-truth landmarks, as well as comparing remote-sensing classified and actual tree canopy sizes of tree individuals was also performed; for further details concerning the remote-sensing methods see (Moustakas et al., 2006; Moustakas et al., 2008).

*Spatio-temporal tree dataset*

The classification conducted populated a database containing the X, Y coordinates of each tree, the year, the canopy surface area in m², a unique tree number ID, the year that the tree was first seen and last seen, and whether the tree survived (survival) or not (death) as binary events. Processing further with spatial statistical analysis, we clustered spatial neighbourhoods in terms of circles with increasing radii spanning from 4, 8, 16, 32, 64, 128, 256, and 512 (focal neighbourhood) m around each tree individual (focal individual) replicated across all tree individuals in the data base, for each available time step. We then calculated (i) the number of tree individuals, (ii) the total canopy cover, and (iii) the percentage of canopy cover within each focal neighbourhood [4, … , 512] m.

*Data integration*

We sought to integrate the spatial tree data set with NDVI, temperature, and soil moisture of neighbouring areas in a grid-based classification of 30 x 30 m. The USGS Landsat 5 Collection 1 Tier 1 Raw Scenes with a resolution of 30 m were processed using a Google Earth Engine (GEE) script (see Supplementary material, Appendix 1). Among the high resolution optical sensors, Landsat-5 is considered as one of the better calibrated sensors for NDVI extraction and it is thus often used as a benchmark for other products (Beck et al., 2011). These data were not available for the two first time snapshots (1940 and 1964) since satellite records became available much later. Nevertheless, they were available for the last

3 snapshots (June 1984, 1993, and 2001, exact dates and scenes shown in Table S1)). Indexes discussed here were used without further calibration since (a) the aggregation level of the satellite image information (30 x 30 m) would not allow direct ground-truthing with conventional ground measurements, and (b) the results we later draw upon don't pertain absolute values but rather a comparative assessment among the scenes. NDVI and LST were extracted from the respective Landsat scene enveloping the study area and values were plotted for each scene pixel (Fig. S1). To get an estimate of the wet and dry edges of the conceptual trapezoid, data was binned in intervals of 0.05 NDVI and for each interval minimum and maximum values are identified. Then, linear regression was applied to the resulting minimum (wet edge) and maximum (dry edge) temperatures. From the regression equations, the slopes (*a*) and intercepts (*b*) were obtained and applied to Eq. 4. Table S1 shows the resulting values (see Supplementary material). The R code for the estimation of SMI from NDVI and LST values is given in Supplementary material, Appendix 2. The final resulting dataset comprised of all the tree data described above plus values of NDVI, temperature, and soil moisture index for each tree individual.

*Standardised Precipitation Index (SPI)*

Standardised Precipitation Index (SPI); (McKee et al., 1993) is widely used to access meteorological drought occurrence as cumulative precipitation deviation from the norm on a variety of timescales (Daliakopoulos et al., 2017). On short timescales, SPI relates well to stress on soil moisture, while at longer timescales, it can depict water stress on slower processes such as groundwater and reservoir storage (Keyantash, 2018). SPI is obtained by fitting a gamma or a Pearson Type III distribution to monthly precipitation values. The default implementation employed here uses a 2-parameter gamma distribution fit where the shape and scale parameters are maximum likelihood estimates as described in (Thom, 1958). Here, SPI-48 (the cumulative precipitation deviation from the norm over 48 months) corresponding to long duration events (Daliakopoulos et al., 2017) was estimated using monthly precipitation data from the nearest (distance ~35 km) weather station at Kimberley, South Africa, and the 'SPI' package in R (R Development Core Team, 2018) for the period 1940-2003. Missing monthly precipitation values $P_{MY}$ (about 13% of the dataset) were infilled based on an unbroken dataset of annual values $P_Y$, considering the monthly average $\overline{P_M}$ for the given month M and the long-term annual average $\overline{P_Y}$ according to:

$$P_{MY} = P_Y \frac{\overline{P_M}}{\overline{P_Y}} \tag{5}$$

SPI-48 allowed us to define periods of long-term water stress or availability that could have an impact on deep-rooted vegetation. SPI values between [-1, 1] consist 68% of the total values, while values between [1.5, 2.0] define severely wet periods and values between [-2.0, -1.5] severely dry periods, while values > 2 or < -2 extremely wet or dry cases, respectively (Keyantash, 2018; McKee et al., 1993).

*Survival Analysis*

We employed Generalised Linear Models with logistic regression with tree death as dependent variable (an event occurring at most once for each tree in the data and thus avoiding temporal autocorrelation). We initially sought to quantify the most parsimonious data-driven index of neighbourhood density which included (a) number of tree individuals within each focal neighbourhood, (b) percentage of canopy cover within each focal neighbourhood, (c) total cover within each focal neighbourhood by selecting the model that exhibited the lowest Akaike (AIC) value (Burnham and Anderson, 2002; Gunton and Kunin, 2007; Moustakas et al., 2018). Sequentially sought to quantify the most parsimonious data-driven index of focal neighbourhood (i.e. define the best scale of competitive-facilitative tree interactions) by selecting the focal neighbourhood (scale) that exhibited the lowest AIC. Having quantified the optimal index of neighbourhood density and the optimal index of spatial scale of interactions, we then sought to quantify tree survival (dependent variable) as a function of the size of tree individual, density, and year and all their two-way and three-way interactions. In particular, the three-way interaction between tree size, density and year can provide spatio-temporal information regarding potential switching from competition to facilitation through time (negative to positive density effects on survival) for different levels of density and how is this modulated by tree size. Year refers to the last seen year of a tree. Analysis of heteroscedasticity of the best model indicated that the model fit assumptions were fulfilled.

*Explaining temperature or SMI with tree size, density, and year*

Linear models were fitted between SMI or LST as dependent variables (analysis repeated twice, once for each dependent variable) and the best data-driven index of density, scale of interactions and tree size (independent variables), in order to predict soil moisture or land surface temperature on each location based on those variables. Analysis of heteroscedasticity of each of the two models, indicated that the model fit assumptions were fulfilled each time.

**Results**

*Data-driven index of density and of spatial scale*

The number of tree individuals increased exponentially across the examined scales of 4 to 512 meters around each tree (Fig. 2a); this pattern was very similar to the total cover around each tree across scales (Fig. 2b). However, the inverse pattern was recorded with the percentage of cover around each tree across scales, where the percentage of cover decreased exponentially across scales (Fig. 2c). The best data-driven index of density was total cover across scales (Fig. 2d and Table S3, S4, S5 in Supplementary material S2). The best data-driven scale of interactions was 512 m, the coarsest scale from the ones explored, followed by 4 m the finest scale from the ones explored (Fig. 2d and Table S4). We therefore proceeded throughout the analysis by counting within a circle of 512 m around each individual tree (scale) the total tree canopy cover in $m^2$ (density).

Examining the best model fit between tree death and density (total cover within 512 m from each tree), size of the individual tree, and year indicated that all three variables were significant and their marginally significant three-way interaction between them was not justified (Table 1); the removal of the three-way interaction between size-density-year resulted in a model with >2 AIC difference than the full model (Burnham and Anderson, 2002). The interaction between year and size explained the largest amount of deviance (696.66) followed by tree size alone (543.27), and density alone (345.41); (Table 1). The interaction between size and density (1.37), size and year (17.77), the three-way interaction between size density and year (7.47) explained relatively small amounts of the total explained model deviance, and year alone was also not among the most informative predictors of survival (125.01); (values of explained deviance in parentheses here and Table 1).

Size had a negative effect on death indicating that larger individuals had lower chance of death and thus a higher chance of survival (negative model coefficient for size alone, Table 2), and this was consistent across years: the size-year interaction was negative across all years with negative coefficients for size:1964, size:1984, and size:1993; Table 2. Increasing densities always resulted in increased chance of survival (facilitation) for small-sized trees across all years (bin size=0, Fig. 3). Increasing densities had no effect on the mean chance of survival for large-sized trees (bin size=350, Fig. 3); however, increased densities resulted in increased the confidence intervals for the survival of big trees indicating larger variance and potentially negative density effects on big trees (bin size=350, Fig. 3). In 1940, the interaction between density and size is negative with increasing densities across all tree sizes (i.e. there was facilitation) except the largest trees (bin size=350) were there was no effect between density and survival (neither facilitation or competition) (Fig. 3). In 1964, there were positive density effects on survival (facilitation) for all tree sizes except the largest trees (bin size=350) where negative density effects were found (competition);  (Fig. 3). In 1984, there were increased deaths with density across all tree sizes (competition) except the smallest sized trees (bin size=0) that the relationship is positive i.e. facilitation (Fig. 3). In 1993 there is decreased chance of death with increasing density (facilitation) for small and intermediate tree sizes and increased death chance with density (competition) for the largest sized trees (bin sizes of 200 and 350); (Fig. 3).

*Water stress (SPI)*

In terms of water stress (SPI index), there were periods of severe draught (SPI < 1.5) as well as severe humidity (SPI > 1.5) across all available time intervals (Fig. 4). Exceptional draught (SPI ≤ -2) was recorded during the 1964 – 1984 period (in 1967; Fig. 4) as well as during the 1984 – 1993 period (in 1986; Fig. 4). Exceptional humidity (SPI ~ 3) was also recorded in the 1964 – 1984 period (in 1978; Fig. 4).

*Soil moisture (SMI)*

Tree size, year, and density were highly significant predictors of SMI (all interactions significant; Table 3). The interaction between year and density and year alone explained the majority of deviance (8.29), but overall the values of deviance explained were low (all values of deviance explained in Table 3 are smaller than 10). SMI increased with tree size, and with canopy cover (positive model coefficients in Table 4). SMI also increased in time during the study as indicated by the positive coefficients for years 1993 and 2001 (the coefficient for year 1984 is in the intercept; Table 4). SMI always decreased with increased densities for large sized trees (Fig. 5). In 1984 SMI increased with density for small sized trees (Fig. 5). In 1983, SMI increased with density for small and middle sized trees (Fig. 5). In 2001 SMI decreased with density across all tree sizes (Fig. 5).

*Land Surface Temperature (LST)*

Tree size, year, and density were highly significant predictors of LST (all covariates and their interactions significant; Table 5). Overall, year explained the vast majority of deviance in LST (deviance explained 62390 in Table 5), followed by tree size (455), and the interaction between density and year (359; Table 5). LST decreased with increasing tree size and density (negative model coefficients in Table 6). LST decreased in 1993 and increased in 2001 in comparison with 1984 (the coefficient for 1984 is within the intercept; Table 6). In 1984 LST decreased with increasing density for small and middle sized trees, while it increased with increasing density for large sized trees (Fig. 6; Table 6). In 1993, LST decreased with density for all tree sizes except the largest trees (bin size=350); (Fig. 6). In 2001, LST increased with density for all tree sizes (Fig. 6).

**Discussion**

Overall, death (the reciprocal of survival) was best explained (in terms of model deviance) by the interaction between density and year, shifting from facilitation to competition through time. It is important to note that it is the same tree individuals that facilitated that end up competing (Wright et al., 2015). Thus, when seen form a dynamic spatio-temporal perspective, in space there are both scale-dependent positive and negative density effects coexisting in the same ecosystem as described in other studies (Riginos et al., 2009; Staver, 2018; van de Koppel et al., 2006) - see also Table S5 & S6 in Supplementary analysis S3 for reporting the same result here. In time there is a shifting of positive to negative density effects at shorter time scales (Wright et al., 2015) as well as through years as shown here. However, one of the major findings reported here is that the survival of small sized trees was always facilitated by increased densities (i.e. always facilitation), while the survival of large sized trees was never facilitated by increased densities (i.e. either competition or no density effects).

It is often considered that within the stress gradient, it is the arid end (i.e. lack of water) that generates stress and thereby promotes facilitation (Dohn et al., 2013; Noemí et al., 2016). However even in an arid ecosystem such as the one examined here there is severe or exceptional water stress from humidity too – see SPI graph, Fig 4. While we cannot causally link water stress temporal conditions with survival and facilitation (we are unaware in which year/point within each interval of the data each tree died), we conclude that experimental stress gradients should include both the arid end as well as the humid end stress. Will facilitation occur in the high end of severe or exceptional humid conditions? Tree death due to prolonged wet conditions is well recorded in humid ecosystems (Assahira et al., 2017; Tzeng et al., 2018) and to that end extreme humidity can act as a stressor. Nevertheless our understanding of the use of surface vs. groundwater by deep rooted trees in more arid ecosystems is limited (Steggles et al., 2017).

The second best explanatory variable of the death was tree size alone while density alone was the third best predictor. During 1940-1964 facilitation was recorded across densities and tree sizes. However, large sized trees exhibited higher variance in the confidence intervals of the positive effects of density, implying that the level of facilitation varied. During 1964-1984 higher densities facilitated small sized trees but resulted in competition for large sized tree individuals. During 1984-1993 higher densities resulted in competition across all tree sizes but competition was higher for larger sized trees. During 1993-2001, density did not show any effects on small sized tress but exhibited negative effects on large sized ones. It is therefore not only the effects of density (Dickie et al., 2005; Wright et al., 2015) but also the size of the individual that play an important role – some earlier work has been briefly mentioning the role of tree size in terms of height in tree-grass interactions (Blaser et al., 2013; Moustakas and Evans, 2013). The role of tree size is important because large trees have a longer rooting system (Jackson et al., 2000) and are thus more likely to up-lift water via hydraulic lift (Ludwig et al., 2003) with vertical roots or to sustain rain water through their horizontal roots (Caldwell et al., 1998; Caldwell and Richards, 1989; Schenk and Jackson, 2002). In addition, tree size is a good predictor of survival (Colangelo et al., 2017; Coomes and Allen, 2007; Moustakas and Evans, 2015) with large trees having a higher probability of survival.

The fact that tree size was more important predictor than density has also serious implications for nature-based solutions (Keesstra et al., 2018; Nesshöver et al., 2017): maintaining large tree individuals or planting species that can become large-sized can safeguard against desertification or tree-less areas (Aba et al., 2017; Lindenmayer and Laurance, 2017; Runnström, 2000) by promoting survival at long time periods through harsh environmental conditions. The role of scattered trees (Prevedello et al., 2018) and, complementarily, the effects of declining old large tress (Jones et al., 2018) on biodiversity conservation have also been highlighted. In addition, species that can grow fast or become large-sized can facilitate ecological restoration of degraded areas (Rawlik et al., 2018). Assuming that large trees are likely to be old (Harper, 1977), in agricultural systems old individuals have been reported to prevent soil erosion to a considerable larger extend than younger ones (Rodrigo-Comino et al., 2018) and to that end old large tress are likely to efficiently act against soil erosion. Large trees are keystone species both in natural ecosystems (Munzbergova and Ward, 2002) as well as in urban parks (Stagoll et al., 2012).

In terms of soil moisture (SMI) the interaction between year and density explained the majority of deviance implying that the effect of density on soil moisture depends on the levels of density. Within this effect, large tree individuals exhibited a negative relationship with SMI for high densities. It has been reported that SMI may be peaking at intermediate densities (Ilstedt et al., 2016). Our results show no support for this but this could be a limitation of the linear assumptions of our analysis (Berk, 2004).

In terms of temperature, LST depended mainly on year meaning that overall it is the physical conditions that define the LST and the role of vegetation is relatively small (the deviance explained by year is two levels of magnitude higher than the one explained by biotic characteristics, size, or density). However, within this effect, both tree size (Breshears et al., 1998) and density (Kawashima, 1994; Song et al., 2013) had an effect in moderating temperature, with tree size having a level of magnitude stronger effect in temperature moderation than density. Again this highlights the importance of large trees on ecosystems (Jones et al., 2018; Lindenmayer and Laurance, 2017). In addition it has implications for nature based solutions in moderating urban temperatures (Gill et al., 2007; Nielsen et al., 2017), water use (Lin et al., 2018), energy saving (Kliman and Comrie, 2004; McPherson and Simpson, 2003; Morakinyo et al., 2018), as well as climatic-efficient agroforestry (Sida et al., 2018). While these effects can be quantified via remote sensing, they may often pass unnoticed as high resolution data are needed (Zhou et al., 2018).

Defining the spatial scale of interactions is critical (Génin et al., 2018) for defining density and this in return can have direct effects on positive or negative density effects. Often the 'local' interactions or fine scale is measured as e.g. distance to the nearest plant neighbours (Li et al., 2017; Meyer et al., 2008). However, the nearest neighbours in an arid ecosystem will be more distant than the nearest neighbours in a humid ecosystem. In addition if for example the four nearest neighbours were measured, the fifth nearest neighbour (not accounted for) may be a large-sized individual with strong interactions with the focal individual (Wang et al., 2018). In designed experimental studies (e.g. planting or manipulating plant individuals (Roush et al., 2017)), setting up quadrats also requires an, often subtly taken, decision regarding the scales of interaction. We suggest that a data-driven definition of scale of local interactions (Gunton and Kunin, 2007) may be a step forward for better understanding positive and negative density effects (Bradter et al., 2013; Gunton and Kunin, 2009), as well as their implications for agriculture (Gunton et al., 2016), soil water availability (Zhang et al., 2018), and potential temperature amelioration (Soliveres et al., 2011; Wright et al., 2015). Note that the results derived here regarding competitive-facilitative interactions and tree survival would be notably different at different spatial scales across the ones examined (Supplementary analysis S3) and the relationship between scale and survival is spatio-temporal (Soliveres et al., 2010) and to that end, complex.

Going a step further, the data-driven scale of interactions (found to be here 512 m), would not hold true when calculated for each available year individually; partitioning the data for each available year and calculating the best data driven scale of interactions in 1940, 1964, 1984, 1993, and 2001 would not yield the same optimal scale of 512 m for each year. However using a dynamic (i.e. changing with time) year-specific spatial scale (circle) of interactions would introduce the statistical problem of multi-collinearity (Arturs, 2018; Fox and Monette, 1992): any scale found to be optimal for at least one time period would need to be included in the model that contains all the data and therefore densities e.g. at 4 m and at 512 m circles would need to be included in a full model across years. However, all individuals at 4 m are also within the 512 m circle generating multi-collinearity (Fox and Monette, 1992). While we chose a 'mean-field-approach' in defining the spatial scale of interactions across the time span of the study, in reality the scale of interactions is shorter than 512 m in some years (results not shown here).

## Conclusions

Defining the spatial scale of interactions has substantial effect on density and reciprocally on whether density interactions will be positive or negative. The data-driven scale of interactions can change between years. Within the best data-driven spatial interaction scale, the best explanatory covariates of tree survival is the interaction between density and year shifting from facilitation to competition through time. Small sized trees are always facilitated by increased densities while large sized trees had either negative or no density effects. Tree size (alone) is a more important predictor than density (alone) in tree survival. This has serious implications for nature-based solutions, as maintaining large tree individuals or planting species that can become large-sized can act against tree-less areas by promoting survival at long time periods through harsh environmental conditions. Large trees have also a significant effect in moderating land surface temperature thereby creating a cool microclimate, and this effect is higher than the one of vegetation density on temperature. Therefore, an equal total cover consisted of several small-sized or middle-sized trees will not moderate the temperature as the same total cover comprised by large-sized trees.

## Acknowledgements

We thank the organizers of the first TerraEnVision conference, Barcelona, 2018 that lead to this special issue. The authors would like to acknowledge Dr. George Azzari from the Centre on Food Security and the Environment, Stanford University, for developing and sharing his thermal analysis GEE script based on Jimenez-Munoz et al. (2009). AM acknowledges a 2017-2018 Conference Grant funding from Universiti Brunei Darussalam.

**References**:

Aba S, Ndukwe O, Amu C, Baiyeri K. The role of trees and plantation agriculture in mitigating global climate change. African Journal of Food, Agriculture, Nutrition and Development 2017; 17: 12691-12707.
Antonio N-CJ, Miguel V, Marta G. Trait-based selection of nurse plants to restore ecosystem functions in mine tailings. Journal of Applied Ecology 2018; 55: 1195-1206.
Arturs K. Multicollinearity: How common factors cause Type 1 errors in multivariate regression. Strategic Management Journal 2018; doi:10.1002/smj.2783.
Assahira C, Piedade MTF, Trumbore SE, Wittmann F, Cintra BBL, Batista ES, et al. Tree mortality of a flood-adapted species in response of hydrographic changes caused by an Amazonian river dam. Forest ecology and management 2017; 396: 113-123.
Baró F, Gómez-Baggethun E. Assessing the Potential of Regulating Ecosystem Services as Nature-Based Solutions in Urban Areas. In: Kabisch N, Korn H, Stadler J, Bonn A, editors. Nature-Based Solutions to Climate Change Adaptation in Urban Areas: Linkages between Science, Policy and Practice. Springer International Publishing, Cham, 2017, pp. 139-158.
Beck HE, McVicar TR, van Dijk AI, Schellekens J, de Jeu RA, Bruijnzeel LA. Global evaluation of four AVHRR–NDVI data sets: Intercomparison and assessment against Landsat imagery. Remote Sensing of Environment 2011; 115: 2547-2563.
Berk RA. Regression analysis: A constructive critique. Vol 11: Sage, 2004.
Bertness MD, Callaway R. Positive interactions in communities. Trends in Ecology & Evolution 1994; 9: 191-193.
Blaser WJ, Sitters J, Hart SP, Edwards PJ, Olde Venterink H. Facilitative or competitive effects of woody plants on understorey vegetation depend on N-fixation, canopy shape and rainfall. Journal of Ecology 2013; 101: 1598-1603.
Bradter U, Kunin WE, Altringham JD, Thom TJ, Benton TG. Identifying appropriate spatial scales of predictors in species distribution models with the random forest algorithm. Methods in Ecology and Evolution 2013; 4: 167-174.
Breshears DD, Nyhan JW, Heil CE, Wilcox BP. Effects of Woody Plants on Microclimate in a Semiarid Woodland:<br/> Soil Temperature and Evaporation in Canopy and Intercanopy Patches. International Journal of Plant Sciences 1998; 159: 1010-1017.
Brooker RW, Maestre FT, Callaway RM, Lortie CL, Cavieres LA, Kunstler G, et al. Facilitation in plant communities: the past, the present, and the future. Journal of Ecology 2008; 96: 18-34.
Burnham KP, Anderson DR. Model Selection and Multimodel Inference. New York: Springer Verlag, 2002.
Caldwell MM, Dawson TE, Richards JH. Hydraulic lift: consequences of water efflux from the roots of plants. Oecologia 1998; 113: 151-161.
Caldwell MM, Richards JH. Hydraulic lift: water efflux from upper roots improves effectiveness of water uptake by deep roots. Oecologia 1989; 79: 1-5.
Callaway RM, King L. Temperature-driven variation in substrate oxygenation and the balance of competition and facilitation. Ecology 1996; 77: 1189-1195.
Cavieres LA, Badano EI, Sierra-Almeida A, Gómez-González S, Molina-Montenegro MA. Positive interactions between alpine plant species and the nurse cushion plant Laretia acaulis do not increase with elevation in the Andes of central Chile. New Phytologist 2006; 169: 59-69.
Chamberlin TC. Studies for students: the method of multiple working hypothesis Journal of Geology 1897; 5: 837-848.
Chen CP, Zhang C-Y. Data-intensive applications, challenges, techniques and technologies: A survey on Big Data. Information Sciences 2014; 275: 314-347.

Choler P, Michalet R, Callaway RM. Facilitation and competition on gradients in alpine plant communities. Ecology 2001; 82: 3295-3308.
Colangelo M, Camarero JJ, Borghetti M, Gazol A, Gentilesca T, Ripullone F. Size Matters a Lot: Drought-Affected Italian Oaks Are Smaller and Show Lower Growth Prior to Tree Death. Frontiers in Plant Science 2017; 8.
Coomes DA, Allen RB. Mortality and tree-size distributions in natural mixed-age forests. Journal of Ecology 2007; 95: 27-40.
Daliakopoulos IN, Grillakis EG, Koutroulis AG, Tsanis IK. Tree Crown Detection on Multispectral VHR Satellite Imagery. Photogrammetric Engineering & Remote Sensing 2009; 75: 1201-1211.
Daliakopoulos IN, Panagea IS, Tsanis IK, Grillakis MG, Koutroulis AG, Hessel R, et al. Yield Response of Mediterranean Rangelands under a Changing Climate. Land Degradation & Development 2017; 28: 1962-1972.
Davis KF, Rulli MC, Garrassino F, Chiarelli D, Seveso A, D'Odorico P. Water limits to closing yield gaps. Advances in Water Resources 2017; 99: 67-75.
Deering D, Haas R. Using LANDSAT digital data for estimating green biomass.[Throckmorton, Texas test site and Great Plains Corridor, US]. 1980.
Deluigi N, Lambiel C, Kanevski M. Data-driven mapping of the potential mountain permafrost distribution. Science of The Total Environment 2017; 590: 370-380.
Deng Y, Wang S, Bai X, Tian Y, Wu L, Xiao J, et al. Relationship among land surface temperature and LUCC, NDVI in typical karst area. Scientific reports 2018; 8: 641.
Dickie IA, Schnitzer SA, Reich PB, Hobbie SE. Spatially disjunct effects of co-occurring competition and facilitation. Ecology Letters 2005; 8: 1191-1200.
Dohn J, Dembélé F, Karembé M, Moustakas A, Amévor KA, Hanan NP. Tree effects on grass growth in savannas: competition, facilitation and the stress-gradient hypothesis. Journal of Ecology 2013; 101: 202-209.
Fayyad UM, Piatetsky-Shapiro G, Smyth P. Knowledge Discovery and Data Mining: Towards a Unifying Framework. KDD. 96, 1996, pp. 82-88.
Fox J, Monette G. Generalized collinearity diagnostics. Journal of the American Statistical Association 1992; 87: 178-183.
Génin A, Majumder S, Sankaran S, Schneider FD, Danet A, Berdugo M, et al. Spatially heterogeneous stressors can alter the performance of indicators of regime shifts. Ecological Indicators 2018.
Gill SE, Handley JF, Ennos AR, Pauleit S. Adapting cities for climate change: the role of the green infrastructure. Built environment 2007; 33: 115-133.
Gunton RM, Firbank LG, Inman A, Winter DM. How scalable is sustainable intensification? Nature Plants 2016; 2: 16065.
Gunton RM, Kunin WE. Density effects at multiple scales in an experimental plant population. Journal of Ecology 2007; 95: 435-445.
Gunton RM, Kunin WE. Density-dependence at multiple scales in experimental and natural plant populations. Journal of Ecology 2009; 97: 567-580.
Gunton RM, Pöyry J. Scale-specific spatial density dependence in parasitoids: a multi-factor meta-analysis. Functional Ecology 2016; 30: 1501-1510.
Gupta A, Santini S, Jain R. In search of information in visual media. Communications of the ACM 1997; 40: 34-42.
Hao X, Li W, Deng H. The oasis effect and summer temperature rise in arid regions-case study in Tarim Basin. Scientific reports 2016; 6: 35418.
Harper JL. Population biology of plants. The University of California: Academic Press, 1977.
Hong H, Tsangaratos P, Ilia I, Liu J, Zhu A-X, Chen W. Application of fuzzy weight of evidence and data mining techniques in construction of flood susceptibility map of Poyang County, China. Science of the Total Environment 2018; 625: 575-588.

Ilstedt U, Bargués Tobella A, Bazié HR, Bayala J, Verbeeten E, Nyberg G, et al. Intermediate tree cover can maximize groundwater recharge in the seasonally dry tropics. Scientific Reports 2016; 6: 21930.

Jackson RB, Sperry JS, Dawson TE. Root water uptake and transport: using physiological processes in global predictions. Trends in Plant Science 2000; 5: 482-488.

Jagadish H, Gehrke J, Labrinidis A, Papakonstantinou Y, Patel JM, Ramakrishnan R, et al. Big data and its technical challenges. Communications of the ACM 2014; 57: 86-94.

Jiménez-Muñoz JC, Cristóbal J, Sobrino JA, Sòria G, Ninyerola M, Pons X. Revision of the single-channel algorithm for land surface temperature retrieval from Landsat thermal-infrared data. IEEE Transactions on geoscience and remote sensing 2009; 47: 339-349.

Jiménez-Muñoz JC, Sobrino JA. A generalized single-channel method for retrieving land surface temperature from remote sensing data. Journal of Geophysical Research: Atmospheres 2003; 108.

Jones GM, Keane JJ, Gutiérrez RJ, Peery MZ. Declining old-forest species as a legacy of large trees lost. Diversity and Distributions 2018; 24: 341-351.

Kawashima S. Relation between vegetation, surface temperature, and surface composition in the tokyo region during winter. Remote Sensing of Environment 1994; 50: 52-60.

Keesstra S, Nunes J, Novara A, Finger D, Avelar D, Kalantari Z, et al. The superior effect of nature based solutions in land management for enhancing ecosystem services. Science of The Total Environment 2018; 610-611: 997-1009.

Kell DB, Oliver SG. Here is the evidence, now what is the hypothesis? The complementary roles of inductive and hypothesis-driven science in the post-genomic era. BioEssays 2004; 26: 99-105.

Kerenyi J, Putsay M. Investigation of land surface temperature algorithms using NOAA AVHRR images. Advances in Space Research 2000; 26: 1077-1080.

Kerr JT, Ostrovsky M. From space to species: ecological applications for remote sensing. Trends in ecology & evolution 2003; 18: 299-305.

Keyantash JNCfARSE. The Climate Data Guide: Standardized Precipitation Index (SPI), Retrieved from https://climatedataguide.ucar.edu/climate-data/standardized-precipitation-index-spi. Last modified 08 Mar 2018, 2018.

Kitchin R. Big Data, new epistemologies and paradigm shifts. Big Data & Society 2014; 1: 2053951714528481.

Kliman SS, Comrie AC. Effects of vegetation on residential energy consumption. Home Energy 2004; 21: 38-42.

Levi L, Cvetkovic V, Destouni G. Data-driven analysis of nutrient inputs and transfers through nested catchments. Science of The Total Environment 2018; 610-611: 482-494.

Li L, Tilman D, Lambers H, Zhang FS. Plant diversity and overyielding: insights from belowground facilitation of intercropping in agriculture. New Phytologist 2014; 203: 63-69.

Li Y, Hui G, Yu S, Luo Y, Yao X, Ye S. Nearest neighbour relationships in Pinus yunnanensis var. tenuifolia forests along the Nanpan River, China. iForest-Biogeosciences and Forestry 2017; 10: 746.

Lillesand T, Kiefer RW, Chipman J. Remote sensing and image interpretation: John Wiley & Sons, 2014.

Lin BB, Egerer MH, Liere H, Jha S, Bichier P, Philpott SM. Local-and landscape-scale land cover affects microclimate and water use in urban gardens. Science of the Total Environment 2018; 610: 570-575.

Lindenmayer DB, Laurance WF. The ecology, distribution, conservation and management of large old trees. Biological Reviews 2017; 92: 1434-1458.

Ludwig F, Dawson TE, de Kroon H, Berendse F, Prins HH. Hydraulic lift in Acacia tortilis trees on an East African savanna. Oecologia 2003; 134: 293-300.

McKee TB, Doesken NJ, Kleist J. The relationship of drought frequency and duration to time scales. Proceedings of the 8th Conference on Applied Climatology. 17. American Meteorological Society Boston, MA, 1993, pp. 179-183.

McPherson EG, Simpson JR. Potential energy savings in buildings by an urban tree planting programme in California. Urban Forestry & Urban Greening 2003; 2: 73-86.

Meyer KM, Ward D, Wiegand K, Moustakas A. Multi-proxy evidence for competition between savanna woody species. Perspectives in Plant Ecology, Evolution and Systematics 2008; 10: 63-72.

Morakinyo TE, Lau KK-L, Ren C, Ng E. Performance of Hong Kong's common trees species for outdoor temperature regulation, thermal comfort and energy saving. Building and Environment 2018; 137: 157-170.

Moran M, Clarke T, Inoue Y, Vidal A. Estimating crop water deficit using the relation between surface-air temperature and spectral vegetation index. Remote sensing of environment 1994; 49: 246-263.

Moschitz H, Roep D, Brunori G, Tisenkopfs T. Learning and innovation networks for sustainable agriculture: processes of co-evolution, joint reflection and facilitation. Taylor & Francis, 2015.

Moustakas A. Fire acting as an increasing spatial autocorrelation force: Implications for pattern formation and ecological facilitation. Ecological Complexity 2015; 21: 142-149.

Moustakas A. Spatio-temporal data mining in ecological and veterinary epidemiology. Stochastic Environmental Research and Risk Assessment 2017; 31: 829–834.

Moustakas A, Evans MR. Integrating Evolution into Ecological Modelling: Accommodating Phenotypic Changes in Agent Based Models. PLoS ONE 2013; 8: e71125.

Moustakas A, Evans MR. Effects of growth rate, size, and light availability on tree survival across life stages: a demographic analysis accounting for missing values and small sample sizes. BMC Ecology 2015; 15: 1-15.

Moustakas A, Guenther M, Wiegand K, Mueller K-H, Ward D, Meyer KM, et al. Long-term mortality patterns of the deep-rooted Acacia erioloba: The middle class shall die! Journal of Vegetation Science 2006; 17: 473-480.

Moustakas A, Katsanevakis S. Editorial: Data Mining and Methods for Early Detection, Horizon Scanning, Modelling, and Risk Assessment of Invasive Species. Frontiers in Applied Mathematics and Statistics 2018; 4: 5.

Moustakas A, Voutsela A, Katsanevakis S. Sampling alien species inside and outside protected areas: Does it matter? Science of The Total Environment 2018; 625: 194-198.

Moustakas A, Wiegand K, Getzin S, Ward D, Meyer KM, Guenther M, et al. Spacing patterns of an Acacia tree in the Kalahari over a 61-year period: How clumped becomes regular and vice versa. Acta Oecologica 2008; 33: 355-364.

Moustakas A, Wiegand K, Meyer KM, Ward D, Sankaran M. perspective: Learning new tricks from old trees: revisiting the savanna question. Frontiers of Biogeography 2010; 2.

Munzbergova Z, Ward D. Acacia trees as keystone species in Negev desert ecosystems. Journal of Vegetation Science 2002; 13: 227-236.

Muro J, Heinmann S, Strauch A, Menz1, 2 G. Land Surface Temperature retrieval in wetlands using Normalized Difference Vegetation Index-emissivity estimation and ASTER emissivity product. Living Planet Symposium. 740, 2016, pp. 153.

Nesshöver C, Assmuth T, Irvine KN, Rusch GM, Waylen KA, Delbaere B, et al. The science, policy and practice of nature-based solutions: An interdisciplinary perspective. Science of The Total Environment 2017; 579: 1215-1227.

Nielsen AB, Hedblom M, Olafsson AS, Wiström B. Spatial configurations of urban forest in different landscape and socio-political contexts: identifying patterns for green infrastructure planning. Urban Ecosystems 2017; 20: 379-392.

Noemí M, Jaime M, Luis P, Sebastián A, A. GL, Tomas S. The sign and magnitude of tree–grass interaction along a global environmental gradient. Global Ecology and Biogeography 2016; 25: 1510-1519.

Platt JR. Strong inference. science 1964; 146: 347-353.

Prevedello JA, Almeida-Gomes M, Lindenmayer DB. The importance of scattered trees for biodiversity conservation: A global meta-analysis. Journal of Applied Ecology 2018; 55: 205-214.

R Development Core Team. R; A language and environment for statistical computing. Vienna, Austria: R Foundation for Statistical Computing, 2018.

Rawlik M, Kasprowicz M, Jagodziński AM, Kaźmierowski C, Łukowiak R, Grzebisz W. Canopy tree species determine herb layer biomass and species composition on a reclaimed mine spoil heap. Science of The Total Environment 2018; 635: 1205-1214.

Reichstein M, Ciais P, Papale D, Valentini R, Running S, Viovy N, et al. Reduction of ecosystem productivity and respiration during the European summer 2003 climate anomaly: a joint flux tower, remote sensing and modelling analysis. Global Change Biology 2007; 13: 634-651.

Riginos C, Grace JB, Augustine DJ, Young TP. Local versus landscape-scale effects of savanna trees on grasses. Journal of Ecology 2009; 97: 1337-1345.

Rodrigo-Comino J, Brevik EC, Cerdà A. The age of vines as a controlling factor of soil erosion processes in Mediterranean vineyards. Science of The Total Environment 2018; 616-617: 1163-1173.

Roush ML, Radosevich SR, Wagner RG, Maxwell BD, Petersen TD. A Comparison of Methods for Measuring Effects of Density and Proportion in Plant Competition Experiments. Weed Science 2017; 37: 268-275.

Runnström MC. Is Northern China Winning the Battle against Desertification? AMBIO: A Journal of the Human Environment 2000; 29: 468-476.

Schenk HJ, Jackson RB. Rooting depths, lateral root spreads and below-ground/above-ground allometries of plants in water-limited ecosystems. Journal of Ecology 2002; 90: 480-494.

Sida TS, Baudron F, Kim H, Giller KE. Climate-smart agroforestry: Faidherbia albida trees buffer wheat against climatic extremes in the Central Rift Valley of Ethiopia. Agricultural and forest meteorology 2018; 248: 339-347.

Smit C, Vandenberghe C, Den Ouden J, Müller-Schärer H. Nurse plants, tree saplings and grazing pressure: changes in facilitation along a biotic environmental gradient. Oecologia 2007; 152: 265-273.

Sobrino JA, Jimenez-Munoz JC, Paolini L. Land surface temperature retrieval from LANDSAT TM 5. Remote Sensing of environment 2004; 90: 434-440.

Sobrino JA, Jiménez-Muñoz JC, Sòria G, Romaguera M, Guanter L, Moreno J, et al. Land surface emissivity retrieval from different VNIR and TIR sensors. IEEE Transactions on Geoscience and Remote Sensing 2008; 46: 316-327.

Soliveres S, DeSoto L, Maestre FT, Olano JM. Spatio-temporal heterogeneity in abiotic factors modulate multiple ontogenetic shifts between competition and facilitation. Perspectives in Plant Ecology, Evolution and Systematics 2010; 12: 227-234.

Soliveres S, Eldridge DJ, Maestre FT, Bowker MA, Tighe M, Escudero A. Microhabitat amelioration and reduced competition among understorey plants as drivers of facilitation across environmental gradients: towards a unifying framework. Perspectives in Plant Ecology, Evolution and Systematics 2011; 13: 247-258.

Song Y, Zhou D, Zhang H, Li G, Jin Y, Li Q. Effects of vegetation height and density on soil temperature variations. Chinese Science Bulletin 2013; 58: 907-912.
Stagoll K, Lindenmayer DB, Knight E, Fischer J, Manning AD. Large trees are keystone structures in urban parks. Conservation Letters 2012; 5: 115-122.
Staver AC. Prediction and scale in savanna ecosystems. New Phytologist 2018; 219: 52-57.
Steggles EK, Holland KL, Chittleborough DJ, Doudle SL, Clarke LJ, Watling JR, et al. The potential for deep groundwater use by Acacia papyrocarpa (Western myall) in a water-limited environment. Ecohydrology 2017; 10: e1791.
Thom HC. A note on the gamma distribution. Monthly Weather Review 1958; 86: 117-122.
Tomlinson CJ, Chapman L, Thornes JE, Baker C. Remote sensing land surface temperature for meteorology and climatology: A review. Meteorological Applications 2011; 18: 296-306.
Tucker CJ. Red and photographic infrared linear combinations for monitoring vegetation. Remote sensing of Environment 1979; 8: 127-150.
Tzeng H-Y, Wang W, Tseng Y-H, Chiu C-A, Kuo C-C, Tsai S-T. Tree mortality in response to typhoon-induced floods and mudslides is determined by tree species, size, and position in a riparian Formosan gum forest in subtropical Taiwan. PloS one 2018; 13: e0190832.
van de Koppel J, Altieri AH, Silliman BR, Bruno JF, Bertness MD. Scale-dependent interactions and community structure on cobble beaches. Ecology Letters 2006; 9: 45-50.
van Helden P. Data-driven hypotheses. EMBO Reports 2013; 14: 104-104.
Veblen KE. Season- and herbivore-dependent competition and facilitation in a semiarid savanna. Ecology 2008; 89: 1532-1540.
Víctor G, Zoraida C, Julian C, Inge A. Facilitation by pioneer shrubs for the ecological restoration of riparian forests in the Central Andes of Colombia. Restoration Ecology 2017; 25: 731-737.
Vlassova L, Perez-Cabello F, Nieto H, Martín P, Riaño D, de la Riva J. Assessment of methods for land surface temperature retrieval from Landsat-5 TM images applicable to multiscale tree-grass ecosystem modeling. Remote Sensing 2014; 6: 4345-4368.
Wang H, Peng H, Hui G, Hu Y, Zhao Z. Large trees are surrounded by more heterospecific neighboring trees in Korean pine broad-leaved natural forests. Scientific Reports 2018; 8: 9149.
Wright A, Schnitzer SA, Dickie IA, Gunderson AR, Pinter GA, Mangan SA, et al. Complex facilitation and competition in a temperate grassland: loss of plant diversity and elevated $CO_2$ have divergent and opposite effects on oak establishment. Oecologia 2013; 171: 449-458.
Wright A, Schnitzer SA, Reich PB. Daily environmental conditions determine the competition–facilitation balance for plant water status. Journal of Ecology 2015; 103: 648-656.
Wright AJ, Wardle DA, Callaway R, Gaxiola A. The Overlooked Role of Facilitation in Biodiversity Experiments. Trends in Ecology & Evolution 2017; 32: 383-390.
Xu C, Holmgren M, Van Nes EH, Maestre FT, Soliveres S, Berdugo M, et al. Can we infer plant facilitation from remote sensing? a test across global drylands. Ecological Applications 2015; 25: 1456-1462.
Yin Z, Feng Q, Wen X, Deo RC, Yang L, Si J, et al. Design and evaluation of SVR, MARS and M5Tree models for 1, 2 and 3-day lead time forecasting of river flow data in a semiarid mountainous catchment. Stochastic Environmental Research and Risk Assessment 2018: doi.org/10.1007/s0047.
Yu K, D'Odorico P. From facilitative to competitive interactions between woody plants and plants with crassulacean acid metabolism (CAM): The role of hydraulic descent. Ecohydrology 2017; 10: e1799.

Zhang Z, Szota C, Fletcher TD, Williams NS, Werdin J, Farrell C. Influence of plant composition and water use strategies on green roof stormwater retention. Science of The Total Environment 2018; 625: 775-781.

Zhou W, Wang J, Qian Y, Pickett STA, Li W, Han L. The rapid but "invisible" changes in urban greenspace: A comparative study of nine Chinese cities. Science of The Total Environment 2018; 627: 1572-1584.

**Table 1.** ANOVA results of a logistic generalised linear model between tree death (dependent variable), and tree size in terms of canopy surface area in $m^2$, tree density in terms of total canopy cover within a circle of 512 $m^2$ around each tree, and year as explanatory variables. Significance codes: 0 '***' 0.001 '**' 0.01 '*' 0.05 '.' 0.1 ' ' 1

| Predictor | Df | Deviance | Residual Df | Residual Deviance | $p(>x)$ | Significance |
|---|---|---|---|---|---|---|
| **None** | | | 10575 | 9846.5 | | |
| size | 1 | 543.27 | 10574 | 9303.2 | <0.0001 | *** |
| s512 | 1 | 345.41 | 10573 | 8957.8 | <0.0001 | *** |
| Year | 3 | 125.01 | 10570 | 8832.8 | <0.0001 | *** |
| size:s512 | 1 | 1.37 | 10569 | 8831.4 | 0.2419 | |
| size:Year | 3 | 17.77 | 10566 | 8813.7 | 0.0005 | *** |
| s512:Year | 3 | 696.66 | 10563 | 8117.0 | <0.0001 | *** |
| size:s512:Year | 3 | 7.47 | 10560 | 8109.5 | 0.05837 | |

**Table 2.** Summary of model coefficients of the ANOVA results from Table 1

| Coefficients | Estimate | Std. Error | z value | $p(>|z|)$ | Significance |
|---|---|---|---|---|---|
| (Intercept) | 3.034 | 30.08 | 10.087 | <0.0001 | *** |
| size | $-2.155 \ 10^{-2}$ | $7.818 \ 10^{-3}$ | -2.756 | 0.0058 | ** |
| s512 | $-4.082 \ 10^{-4}$ | $4.031 \ 10^{-5}$ | -10.127 | <0.0001 | *** |
| Year1964 | $-6.283E \ 10^{-1}$ | $3.820 \ 10^{-1}$ | -1.645 | 0.01 | |
| Year1984 | -3.663 | $3.244 \ 10^{-1}$ | -11.291 | <0.0001 | *** |
| Year1993 | -3.444 | $3.272 \ 10^{-1}$ | -10.525 | <0.0001 | *** |
| size:s512 | $1.952 \ 10^{-7}$ | $1.113 \ 10^{-6}$ | 0.175 | 0.8608 | |
| size:Year1964 | $-1.443 \ 10^{-2}$ | $1.023 \ 10^{-2}$ | -1.411 | 0.1582 | |
| size:Year1984 | $-3.516 \ 10^{-2}$ | $1.004 \ 10^{-2}$ | -3.502 | 0.0005 | *** |
| size:Year1993 | $-6.734 \ 10^{-3}$ | $8.906 \ 10^{-3}$ | -0.756 | 0.4496 | |
| s512:Year1964 | $-4.790 \ 10^{-5}$ | $5.283 \ 10^{-5}$ | -0.907 | 0.3646 | |
| s512:Year1984 | $3.822 \ 10^{-4}$ | $4.098 \ 10^{-5}$ | 9.326 | <0.0001 | *** |
| s512:Year1993 | $3.572 \ 10^{-4}$ | $4.114 \ 10^{-5}$ | 8.683 | <0.0001 | *** |
| size:s512:Year1964 | $1.822 \ 10^{-6}$ | $1.388 \ 10^{-6}$ | 1.312 | 0.1893 | |
| size:s512:Year1984 | $1.421 \ 10^{-6}$ | $1.160 \ 10^{-6}$ | 1.226 | 0.2203 | |
| size:s512:Year1993 | $4.155 \ 10^{-7}$ | $1.144 \ 10^{-6}$ | 0.363 | 0.7164 | |

**Table 3.** ANOVA results of a generalised linear model between Soil Moisture Index (SMI; dependent variable), and tree size in terms of canopy surface area in m$^2$, tree density in terms of total canopy cover within a circle of 512 m around each tree, and year as explanatory variables. Significance codes: 0 '***' 0.001 '**' 0.01 '*' 0.05 '.' 0.1 ' ' 1

| Predictor | Df | Deviance | Residual Df | Residual Deviance | $p(>x)$ | Significance |
|---|---|---|---|---|---|---|
| **None** | | | 11875 | 50.105 | | |
| **size** | 1 | 0.0008 | 11874 | 50.104 | 0.6239 | |
| **s512** | 1 | 0.0065 | 11873 | 50.097 | 0.1563 | |
| **Year** | 2 | 2.9051 | 11871 | 47.192 | <0.0001 | *** |
| **size:s512** | 1 | 0.2540 | 11870 | 46.938 | <0.0001 | *** |
| **size:Year** | 2 | 0.3746 | 11868 | 46.564 | <0.0001 | *** |
| **s512:Year** | 2 | 8.2869 | 11866 | 38.277 | <0.0001 | *** |
| **size:s512:Year** | 2 | 0.1417 | 11864 | 38.135 | <0.0001 | *** |

**Table 4.** Summary of model coefficients of the ANOVA results from Table 3

| Coefficients | Estimate | Std. Error | z value | $p(>|z|)$ | Significance |
|---|---|---|---|---|---|
| **(Intercept)** | 2.667 10$^{-1}$ | 2.394 10$^{-3}$ | 111.372 | <0.0001 | *** |
| **size** | 6.827 10$^{-4}$ | 4.878 10$^{-5}$ | 13.996 | <0.0001 | *** |
| **s512** | 4.956 10$^{-6}$ | 1.499 10$^{-5}$ | 33.064 | <0.0001 | *** |
| **Year1993** | 2.285 10$^{-2}$ | 3.546 10$^{-3}$ | 6.445 | <0.0001 | *** |
| **Year2001** | 1.378 10$^{-1}$ | 3.345 10$^{-3}$ | 41.322 | <0.0001 | *** |
| **size:s512** | -3.246 10$^{-8}$ | 3.307 10$^{-9}$ | -9.815 | <0.0001 | *** |
| **size:Year1993** | -6.327 10$^{-4}$ | 6.862 10$^{-5}$ | -9.220 | <0.0001 | *** |
| **size:Year2001** | -6.619 10$^{-4}$ | 7.521 10$^{-5}$ | -8.800 | <0.0001 | *** |
| **s512:Year1993** | -3.202 10$^{-6}$ | 2.124 10$^{-7}$ | -15.079 | <0.0001 | *** |
| **s512:Year2001** | -8.411 10$^{-6}$ | 2.061 10$^{-7}$ | -40.819 | <0.0001 | *** |
| **size:s512:Year1993** | 2.625 10$^{-8}$ | 4.413 10$^{-9}$ | 5.950 | <0.0001 | *** |
| **size:s512:Year1993** | 2.689 10$^{-8}$ | 4.809 10$^{-9}$ | 5.591 | <0.0001 | *** |

**Table 5.** ANOVA results of a generalised linear model between Land Surface Temperature (LST; dependent variable), and tree size in terms of canopy surface area in $m^2$, tree density in terms of total canopy cover within a circle of 512 m around each tree, and year as explanatory variables. Significance codes: 0 '***' 0.001 '**' 0.01 '*' 0.05 '.' 0.1 ' ' 1

| Predictor | Df | Deviance | Residual Df | Residual Deviance | $p(>x)$ | Significance |
|---|---|---|---|---|---|---|
| size | 1 | 455 | 11874 | 65212 | < 0.0001 | *** |
| s512 | 1 | 37 | 11873 | 65176 | < 0.0001 | *** |
| Year | 2 | 62390 | 11871 | 2785 | < 0.0001 | *** |
| size:s512 | 1 | 19 | 11870 | 2766 | < 0.0001 | *** |
| size:Year | 2 | 23 | 11868 | 2743 | < 0.0001 | *** |
| s512:Year | 2 | 359 | 11866 | 2384 | < 0.0001 | *** |
| size:s512:Year | 2 | 10 | 11864 | 2373 | < 0.0001 | *** |

**Table 6.** Summary of model coefficients of the ANOVA results from Table 5

| Coefficients | Estimate | Std. Error | z value | $p(>|t|)$ | Significance |
|---|---|---|---|---|---|
| (Intercept) | $2.864 \cdot 10^2$ | $1.889 \cdot 10^{-2}$ | 15161.547 | < 0.0001 | *** |
| size | $-5.724 \cdot 10^{-3}$ | $3.848 \cdot 10^{-4}$ | -14.876 | < 0.0001 | *** |
| s512 | $-3.980 \cdot 10^{-5}$ | $1.182 \cdot 10^{-6}$ | -33.663 | < 0.0001 | *** |
| Year1993 | -2.934 | $2.979 \cdot 10^{-2}$ | -104.883 | < 0.0001 | *** |
| Year2001 | 2.130 | $2.631 \cdot 10^{-2}$ | 80.952 | < 0.0001 | *** |
| size:s512 | $2.812 \cdot 10^{-7}$ | $2.609 \cdot 10^{-8}$ | 10.778 | < 0.0001 | *** |
| size:Year1993 | $5.262 \cdot 10^{-3}$ | $5.414 \cdot 10^{-4}$ | 9.719 | < 0.0001 | *** |
| size:Year2001 | $5.370 \cdot 10^{-3}$ | $5.933 \cdot 10^{-4}$ | 9.051 | < 0.0001 | *** |
| s512:Year1993 | $1.901 \cdot 10^{-5}$ | $1.676 \cdot 10^{-6}$ | 11.347 | < 0.0001 | *** |
| s512:Year2001 | $5.698 \cdot 10^{-5}$ | $1.626 \cdot 10^{-6}$ | 34.434 | < 0.0001 | *** |
| size:s512:Year1993 | $-2.144 \cdot 10^{-7}$ | $3.481 \cdot 10^{-8}$ | -6.160 | < 0.0001 | *** |
| size:s512:Year2001 | $-2.362 \cdot 10^{-7}$ | $3.794 \cdot 10^{-8}$ | -6.226 | < 0.0001 | *** |

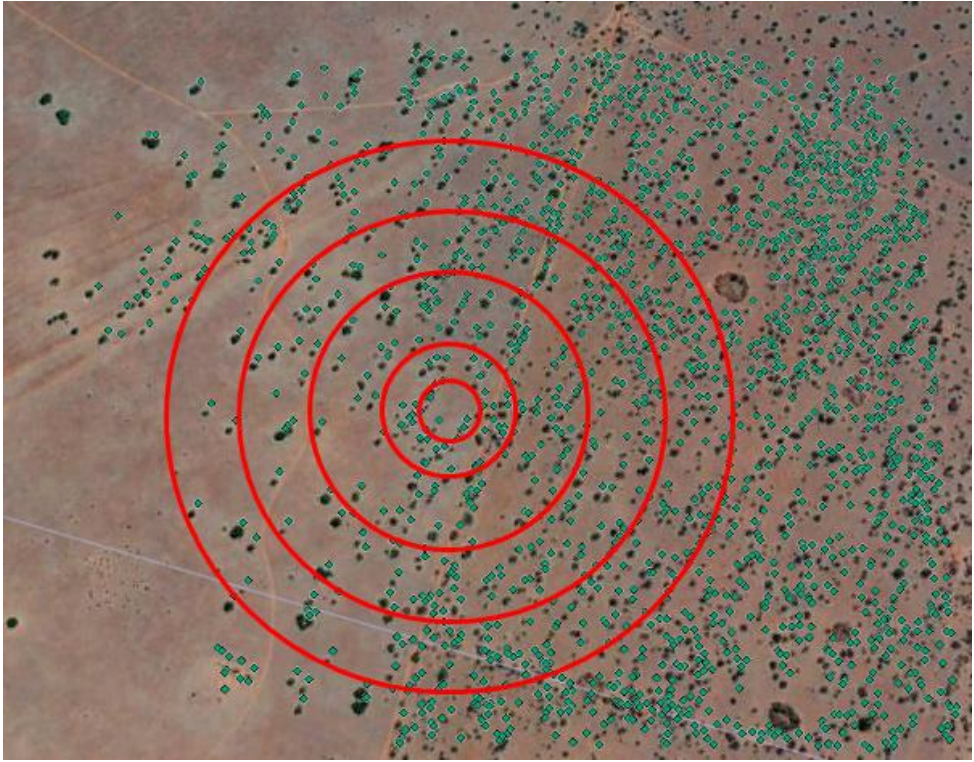

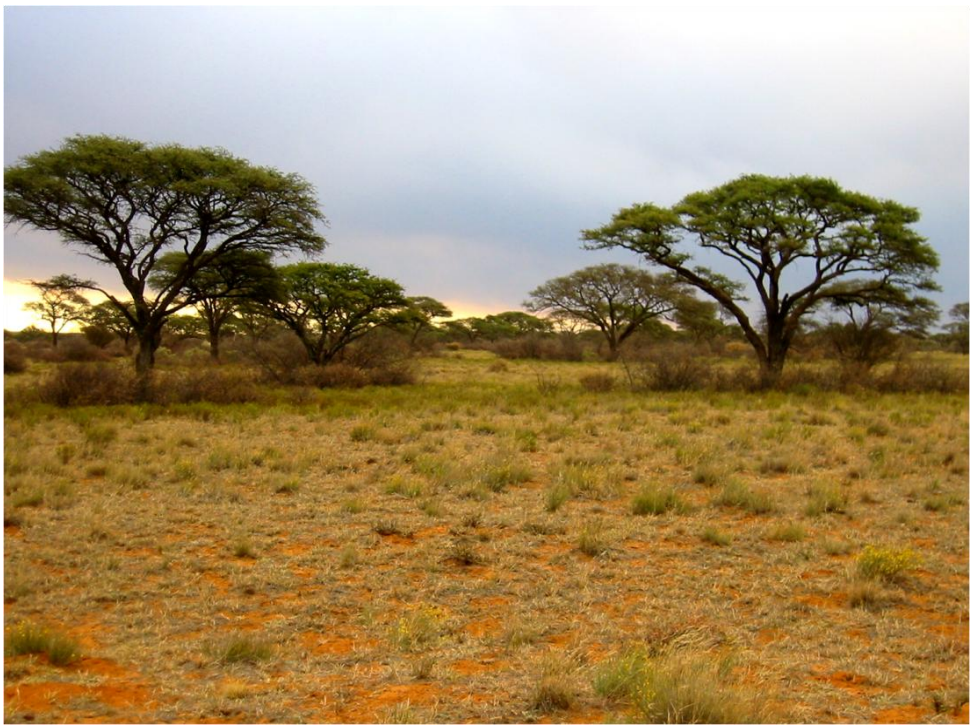

**Figure 1. a.** Examples of potential interaction scales around a tree individual that can be used to define tree densities, and sequentially positive or negative density effects, from a detail of the remote sensing imagery in the study plots. The spatial extent of neighbourhood is used to define density as it is the denominator of the number of individuals or cover within the defined space. This is critical because based on the definition of the scale of interactions (neighbourhood) facilitation can be recorded in one scale while competition can be recorded on the nearest available used. In other words what is the local scale of tree-tree interactions? In addition, in order to investigate the potential existence of local scale

facilitation and landscape scale competition one needs to define what is 'local' and what is landscape' as these often derive from the availability and scale of the data used and the terms are arbitrary. More often than not scales of interactions are taken as a silent presupposition. In this example densities are considerably higher (more trees and higher canopy cover) at finer spatial scales. b. A detail of the study plots as seen from the ground (photo A. Moustakas).

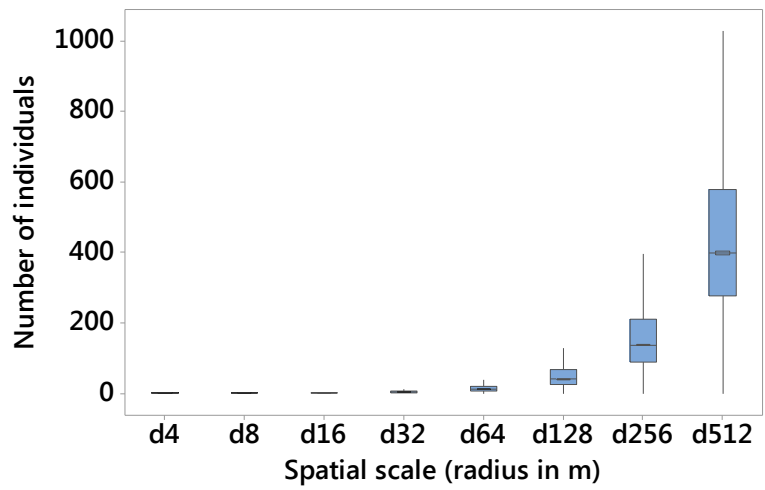

a

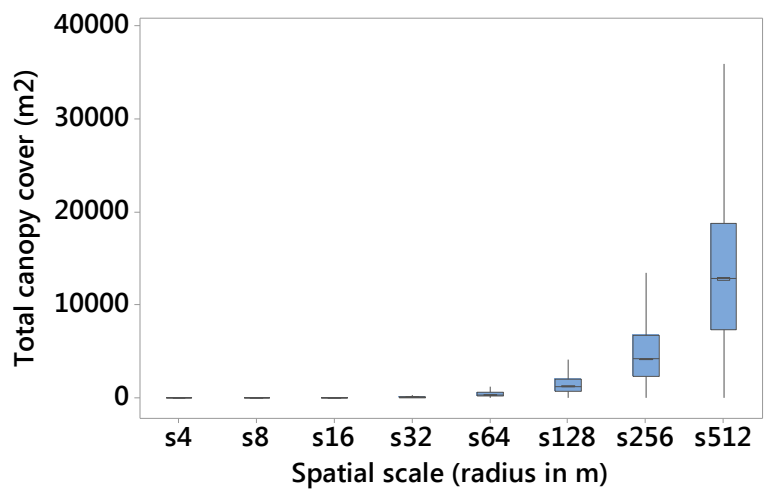

b

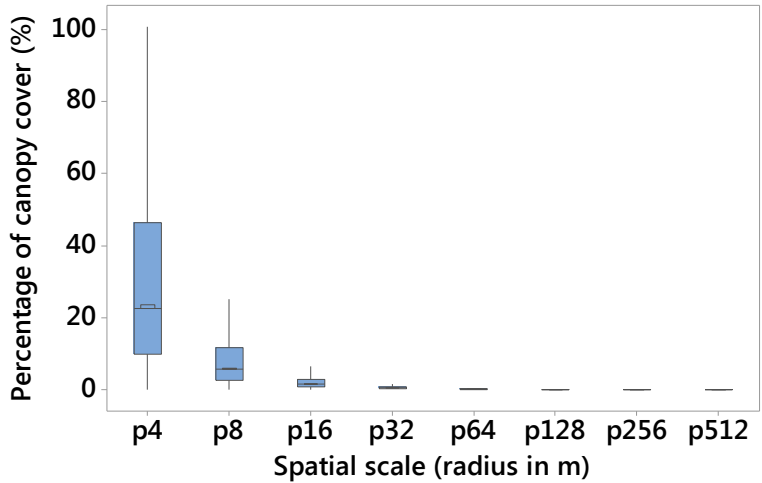

c

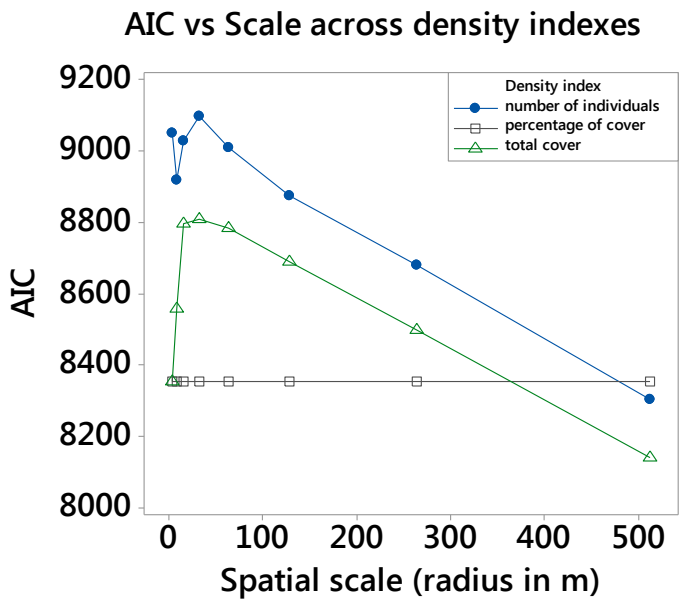

d

**Figure 2.** Indices of neighbourhood density across spatial scales in terms of boxplots: The solid line is the median, and the boxes are defined by the upper and lower quartile (25th and 75th percentiles). The whiskers extend up to 1.5 times the inter-quartile range of the data. Spatial scales include circles with radii of 4, 8, 16, 32, 64, 128, 256, 512 meters around each tree individual. **(a)** Number of tree individuals within each circle. **(b)** Total tree canopy in m$^2$ within each circle. **(c)** Percentage of tree canopy cover (%) within each circle. (d) AIC values of the statistical models explaining survival as a function of tree size, density, year, and their 2-way and 3-way interactions across scales, and indices of neighbourhood density. Neighbourhood density indices included number of trees within each scale, total canopy surface area cover, and percentage of canopy surface area cover within each scale. The statistical models fitted were 24 = 8 scales x 3 indices of neighbourhood. The most parsimonious model (lowest AIC) included total canopy cover as an indicator of neighbourhood and a spatial scale of a circle of 512 m around each tree individual.

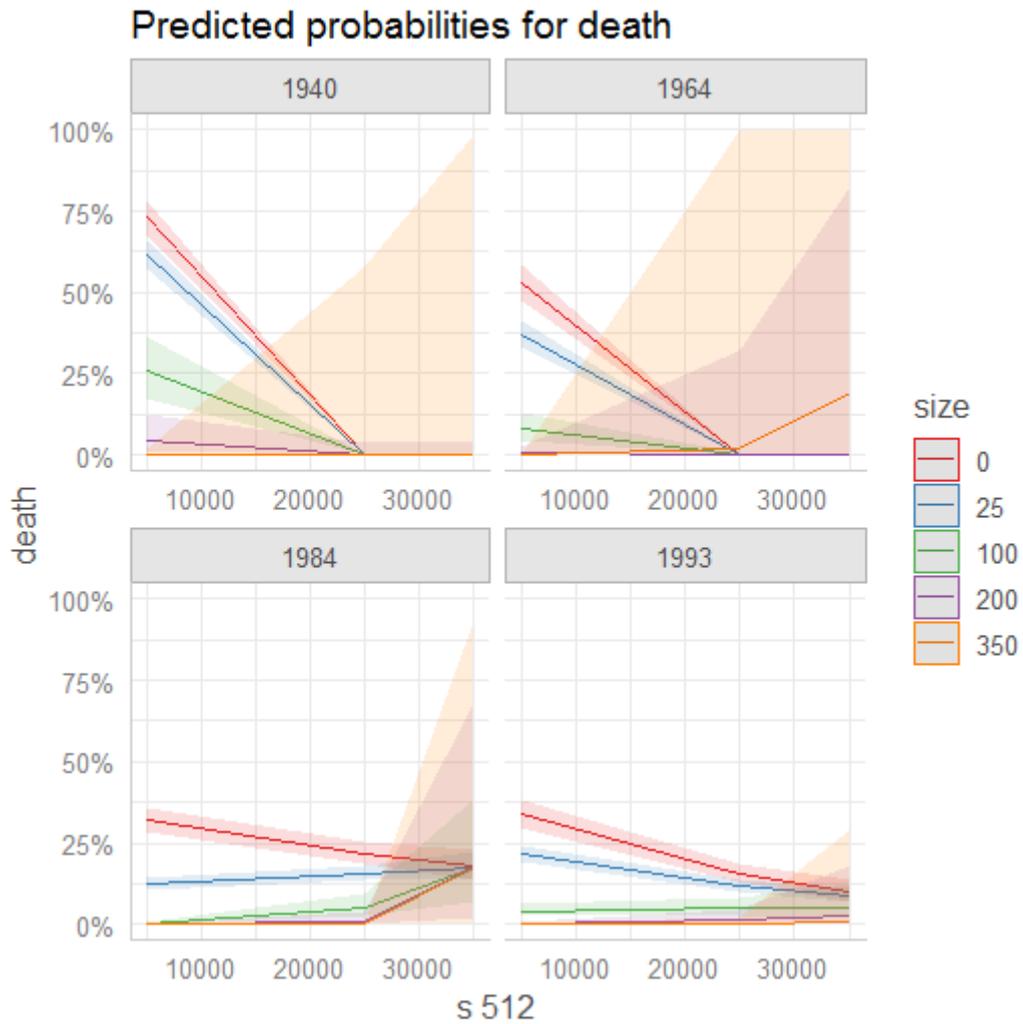

**Figure 3.** Three-way interactions between tree death explained by year, tree size ($m^2$), and density in 512 m circles around each tree. The vertical axis indicates probability of tree death (%). The horizontal axis indicates density in terms of total canopy cover in $m^2$ within a circle of 512 m around each tree. Panels indicate different years. Lines within panels indicate levels of the size of the tree individual that died across years with confidence intervals. Results are also reported in Table 1 and Table 2.

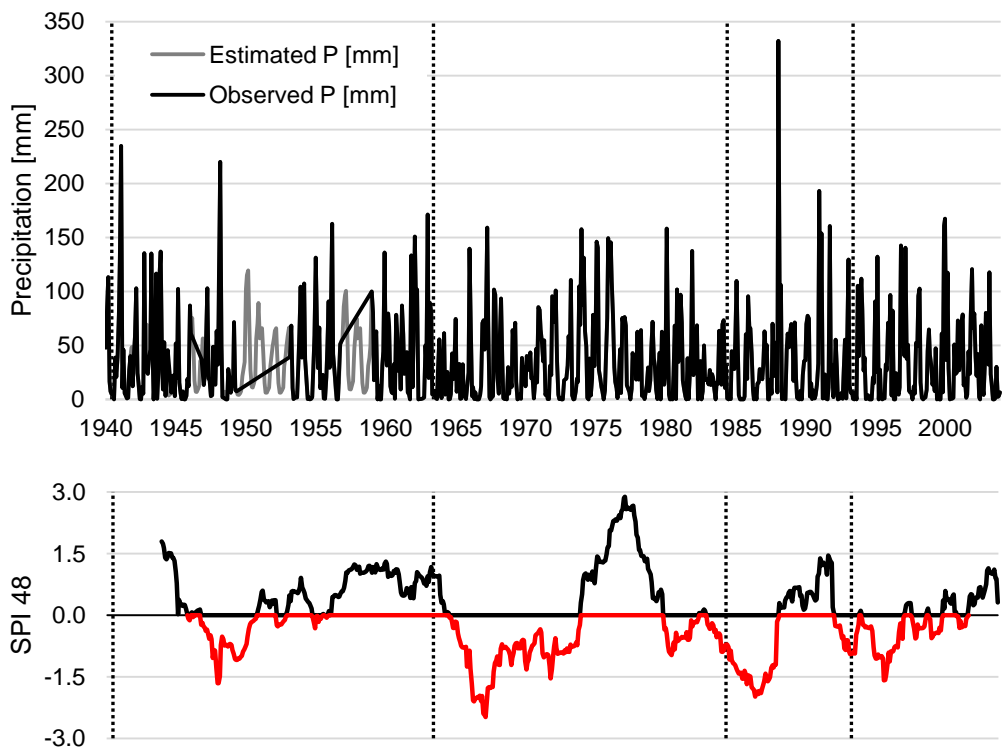

**Figure 4.** Monthly precipitation (in mm; upper panel) and the corresponding Standardised Precipitation Index (SPI; lower panel). Monthly precipitation values derive from the nearest available weather station in Kimberley, South Africa ranging from January 1940 to December 2003. Missing monthly precipitation values (plotted with grey colour on the upper panel) were interpolated (see methods for details). The SPI is calculated from the monthly precipitation time series, it defines periods of humidity and drought, and SPI values are universally comparable. Negative SPI values indicate draught (plotted in red) while positive indicate humidity (plotted in black). Values close to ±1.5 indicate severe conditions while values ±2 indicate exceptional conditions. Grey dotted vertical lines indicate the year when the aerial photos/satellite images are available.

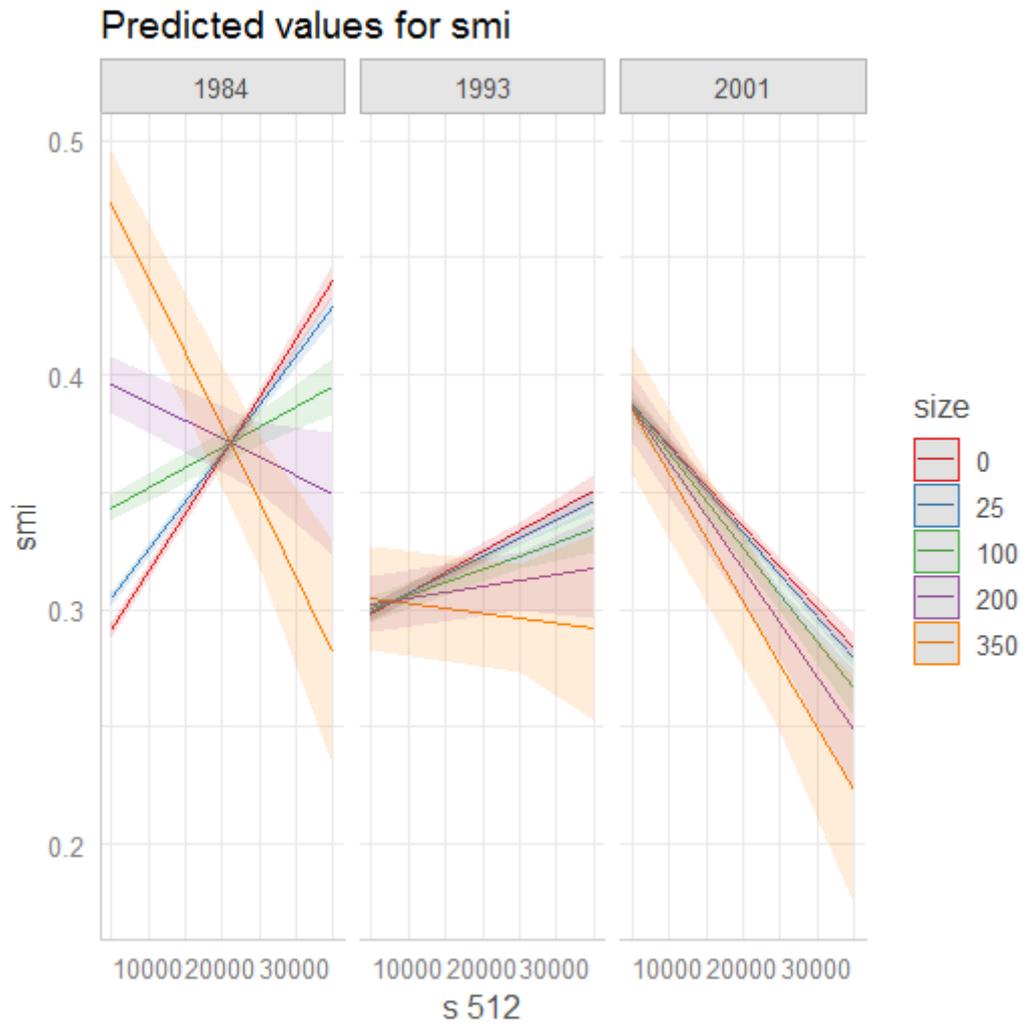

**Figure 5.** Three-way interactions between Soil Moisture Index (SMI) explained by year, tree size (m$^2$), and density in 512 m circles around each tree. The vertical axis indicates SMI values. The horizontal axis indicates density in terms of total canopy cover in m$^2$ within a circle of 512 m around each tree. Panels indicate different years. Lines within panels indicate levels of the size of the tree individual that died across years with confidence intervals. Results are also reported in Table 3 and Table 4.

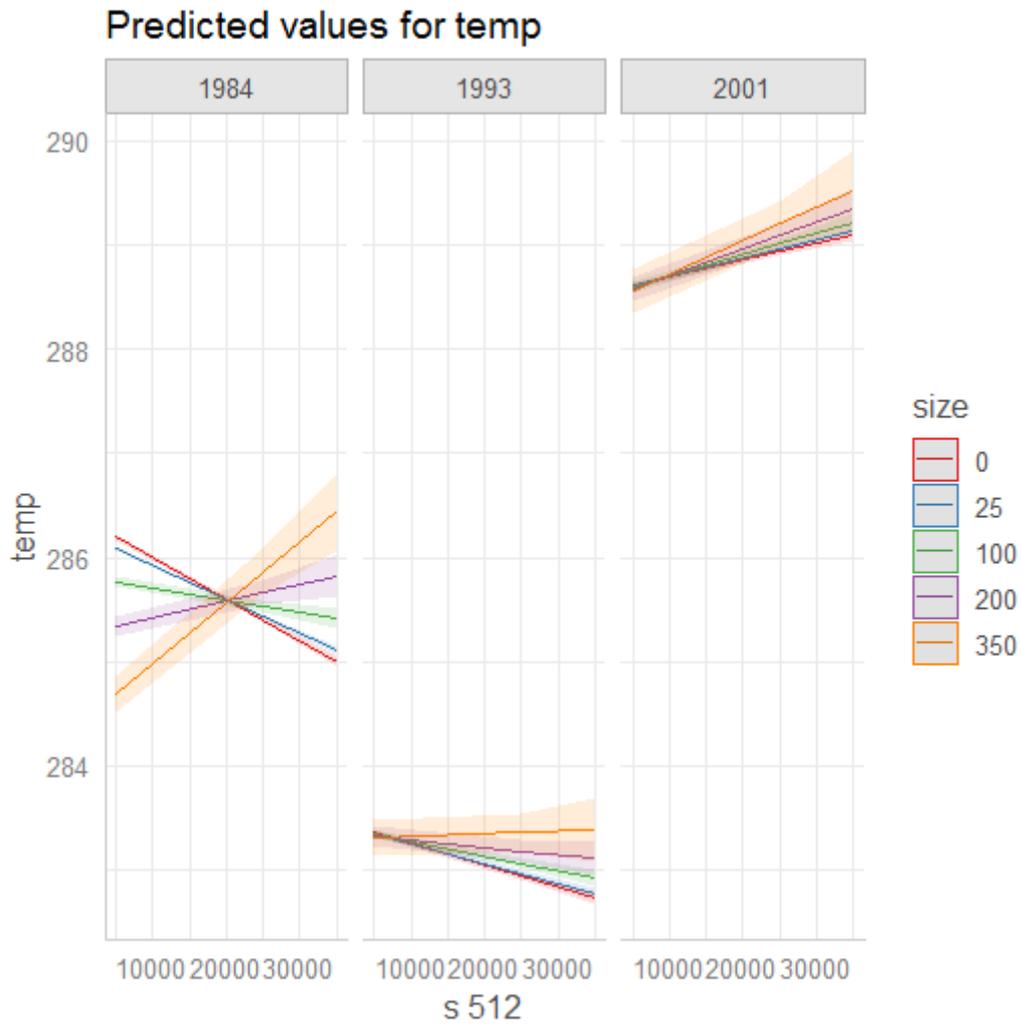

**Figure 6.** Three-way interactions between Land Surface Temperature (LST) explained by year, tree size (m$^2$), and density in 512 m circles around each tree. The vertical axis indicates LST in K$^0$ values. The horizontal axis indicates density in terms of total canopy cover in m$^2$ within a circle of 512 m around each tree. Panels indicate different years. Lines within panels indicate levels of the size of the tree individual that died across years with confidence intervals. Results are also reported in Table 5 and Table 6.

# Supplementary material

# Data-driven competitive-facilitative interactions and their implications for nature based solutions

Supplementary information regarding the Land Surface Temperature (LST) – see methods.

**Table S1: Slopes ($a$) and intercepts ($b$) for the regression equations of Land Surface Temperature of Figure S1.**

| Dataset | | Slope [°K] | Intercept [°K] |
|---|---|---|---|
| June 20, 1984 | Minimum | 5.39 | 279.35 |
| LT05_L1TP_172080_19840620_20170220_01_T1 | Maximum | -7.04 | 290.15 |
| June 13, 1993 | Minimum | 13.20 | 272.8 |
| LT05_L1TP_172080_19930613_20170118_01_T1 | Maximum | -9.61 | 288.65 |
| June 3, 2001 | Minimum | 4.24 | 283.73 |
| LT05_L1TP_172080_20010603_20161210_01_T1 | Maximum | -5.94 | 292.57 |

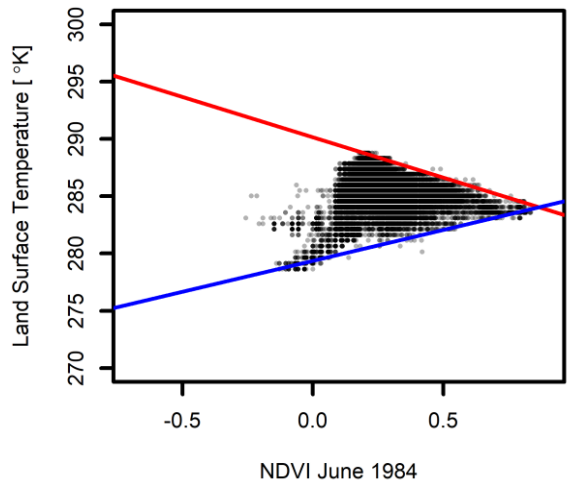

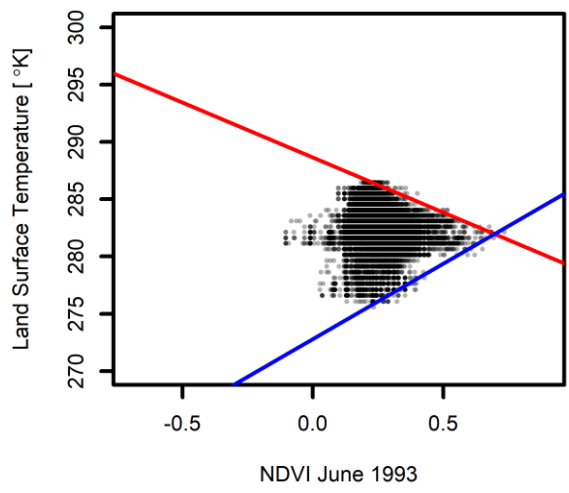

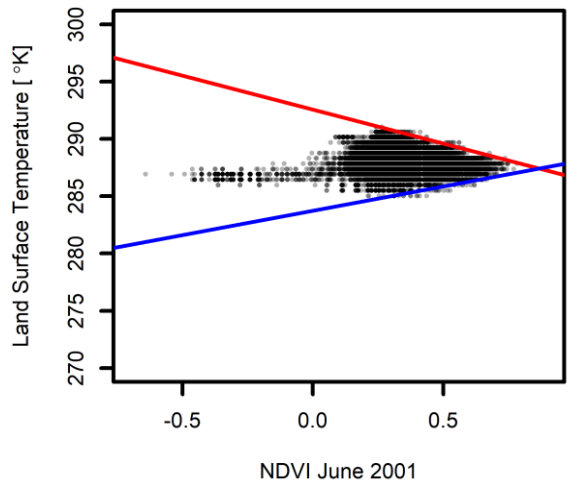

**Figure S1:** NDVI versus Land Surface Temperature [°K] scatterplots (black dots) for June 1984, 1993, and 2001, and respective linear regression lines to identify dry edge (red line) and wet edge (blue line).

# Supplementary analysis S2

Selecting of the best data-driven index of neighbourhood density and the optimal scale of interactions. The potential neighbourhood density indices explored were (i) number of tree individuals denoted with $d$ in the model structure below, (ii) the total tree canopy cover denoted $s$ in the model structure below, and (iii) the percentage of tree canopy cover denoted with $p$ in the model structure below. Each density index was tested across spatial scales of 4, 8, 16, 32, 64, 128, 264, 512 meters (focal neighbourhoods). Years examined are 1940, 1964, 1984, and 1993. Year refers to the last seen year of a tree (i.e. if a tree was seen in 1940 but not in 1964 death year is 1940). Tree deaths were analyzed (the reciprocal of survival) an event occurring maximum once in the dataset as time series. The Generalised linear model (glm) with a binomial family (logistic regression) quantified tree deaths as a function of density within in scale-specific neighbourhood, size in terms of canopy surface area in $m^2$ of the tree individual, and year. The notation * between two variables A* B denotes the effects of variable A, the effects of variable B, and the interaction effect between A and B.

## Number of individual trees across scales

```
>
> q1<-glm(death~size*Year*d4, family="binomial")
> q2<-glm(death~size*Year*d8, family="binomial")
> q3<-glm(death~size*Year*d16, family="binomial")
> q4<-glm(death~size*Year*d32, family="binomial")
> q5<-glm(death~size*Year*d64, family="binomial")
> q6<-glm(death~size*Year*d128, family="binomial")
> q7<-glm(death~size*Year*d256, family="binomial")
> q8<-glm(death~size*Year*d512, family="binomial")
>
> AIC(q1,q2,q3,q4,q5,q6,q7,q8)
```

**Table S2.** AIC scores of logistic regression with number of individual trees as density indicator across scales

```
   df      AIC
q1 16 9050.053
q2 16 8919.362
q3 16 9027.675
q4 16 9096.016
q5 16 9008.763
q6 16 8874.196
q7 16 8679.175
q8 16 8302.611
```

## Total canopy cover across scales

```
a1<-glm(death~size*Year*s4, family="binomial")
a2<-glm(death~size*Year*s8, family="binomial")
a3<-glm(death~size*Year*s16, family="binomial")
a4<-glm(death~size*Year*s32, family="binomial")
a5<-glm(death~size*Year*s64, family="binomial")
a6<-glm(death~size*Year*s128, family="binomial")
a7<-glm(death~size*Year*s256, family="binomial")
a8<-glm(death~size*Year*s512, family="binomial")

> AIC(a1,a2,a3,a4,a5,a6,a7,a8)
```

**Table S3.** AIC scores of logistic regression with total tree canopy cover as density indicator across scales. The bolded italicised value corresponds to the minimum recorded AIC score derived from deaths at scales of 512 m with total canopy cover as a neighbourhood index.

```
   df      AIC
a1 16 8353.873
a2 16 8558.490
a3 16 8796.717
a4 16 8809.134
a5 16 8784.543
a6 16 8688.365
a7 16 8497.747
```
*a8 16 8141.533*

**Percentage of canopy cover across scales**

```
> v1<-glm(death~size*Year*p4, family="binomial")
> v2<-glm(death~size*Year*p8, family="binomial")
> v3<-glm(death~size*Year*p16, family="binomial")
> v4<-glm(death~size*Year*p32, family="binomial")
> v5<-glm(death~size*Year*p64, family="binomial")
> v6<-glm(death~size*Year*p128, family="binomial")
> v7<-glm(death~size*Year*p256, family="binomial")
> v8<-glm(death~size*Year*p512, family="binomial")
>
> AIC(v1,v2,v3,v4,v5,v6,v7,v8)
```

**Table S4.** AIC scores of logistic regression with percentage of canopy cover as density indicator across scales.

```
df        AIC
v1 16 8353.873
v2 16 8353.873
v3 16 8353.873
v4 16 8353.873
v5 16 8353.873
v6 16 8353.873
v7 16 8353.873
v8 16 8353.871
>
```

# Supplementary analysis S3

Results from the second best (data-driven) optimal scale of interactions. Basel on the results of Table S3 the optimal spatial scale of interactions is a circle of 512 m around each tree, which was used throughout the analysis. The second best scale as deduced from results in Table S3 is the one of 4 m (the finest scale from the ones examined here). The results from the scale of 4 m do not match the ones of 512 m and in many years there is an inverse result regarding the effects of density on survival (death), showing that both positive and negative effects coexist on the same location at the same time depending on the scale.

**Table S5.** ANOVA results of a logistic generalised linear model between tree death (dependent variable), and tree size in terms of canopy surface area in $m^2$, tree density in terms of total canopy cover within a circle of 4 $m^2$ around each tree, and year as explanatory variables.

```
              Df Deviance Resid. Df Resid. Dev  Pr(>Chi)
NULL                         10575     9846.5
size           1   543.27    10574     9303.2 < 2.2e-16 ***
s4             1   274.92    10573     9028.3 < 2.2e-16 ***
Year           3   251.50    10570     8776.8 < 2.2e-16 ***
size:s4        1   399.49    10569     8377.3 < 2.2e-16 ***
size:Year      3    19.00    10566     8358.3 0.0002733 ***
s4:Year        3    19.88    10563     8338.4 0.0001794 ***
size:s4:Year   3    16.54    10560     8321.9 0.0008767 ***
---
Signif. codes:  0 '***' 0.001 '**' 0.01 '*' 0.05 '.' 0.1 ' ' 1
```

**Table S6.** Summary of model coefficients of the ANOVA results from Table S5

```
Coefficients:
                   Estimate Std. Error z value Pr(>|z|)
(Intercept)       6.737e-01  1.214e-01   5.551 2.84e-08 ***
size             -1.352e-02  3.885e-03  -3.479 0.000504 ***
s4               -4.227e-02  4.012e-03 -10.537  < 2e-16 ***
Year1964         -6.042e-01  1.670e-01  -3.618 0.000297 ***
Year1984         -1.028e+00  1.419e-01  -7.240 4.49e-13 ***
Year1993         -1.235e+00  1.453e-01  -8.505  < 2e-16 ***
size:s4           2.440e-04  3.440e-05   7.093 1.31e-12 ***
size:Year1964    -7.441e-03  5.903e-03  -1.261 0.207469
size:Year1984    -7.894e-03  5.231e-03  -1.509 0.131231
size:Year1993    -1.146e-02  5.032e-03  -2.277 0.022773 *
s4:Year1964      -2.414e-02  7.022e-03  -3.438 0.000585 ***
s4:Year1984      -1.174e-02  6.005e-03  -1.955 0.050592 .
s4:Year1993       1.198e-02  5.612e-03   2.134 0.032810 *
size:s4:Year1964  1.520e-04  5.173e-05   2.939 0.003297 **
size:s4:Year1984  2.826e-05  4.140e-05   0.683 0.494867
size:s4:Year1993 -1.467e-05  3.972e-05  -0.369 0.711948
```

# Appendix 1

```
// Thermal analysis
// Author: George Azzari
// Center on Food Security and the Environment
// Department of Earth System Science
// Stanford University

/* Based on: Jimenez-Munoz, J.C.; Cristobal, J.; Sobrino, J.A.; Soria, G.; Ninyerola, M.;
Pons, X.; Pons, X.,
        "Revision of the Single-Channel Algorithm for Land Surface Temperature Retrieval
        From Landsat Thermal-Infrared Data,"
        Geoscience and Remote Sensing,
        IEEE Transactions on , vol.47, no.1, pp.339,349, Jan. 2009
        doi: 10.1109/TGRS.2008.2007125 */

//-------------------------------------------------------------------------------------------------
//---------------------------------------Joining Collections---------------------------------------
function filterCollection(imgcoll, start_date, end_date, poly){
  return imgcoll.filterDate(start_date, end_date)
          .filterBounds(poly.centroid()); //using the polygon only would go bananas
          // .filter(ee.Filter.lt('CLOUD_COVER', 10));
}

// Return a Landsat 5 calibrated radiance collection with only thermal (radiative temp).
// Collection is filtered by given dates and by given polygon.
function getLandsatRAD(startdate, enddate, poly){
  var radcoll = filterCollection(ee.ImageCollection("LT5_L1T"), startdate, enddate, poly);
  return radcoll.map(function(img){
      return ee.Algorithms.Landsat.calibratedRadiance(img)
          .select(['B6'], ['B6_RAD'])
          .set({'system:time_start':img.get('system:time_start')});
          });
}

// Return a Landsat 5 TOA collection with only thermal (brightness temp)
// and cloud score band. Collection is filtered by given dates and
// by given polygon.
function getLandsatTOA(startdate, enddate, poly){
  var l5toas = filterCollection(ee.ImageCollection('LANDSAT/LT5_L1T_TOA'), startdate, enddate, poly)
          .map(ee.Algorithms.Landsat.simpleCloudScore)
          .select([5,7], ["B6_BRT","CLOUDSC"]);
  return ee.ImageCollection(l5toas);
}

//Return a Landsat 5 SR collection of surface reflectance and
//quality bands. Collection is filtered by given dates and
//by given polygon.
function getLandsatSR(startdate, enddate, poly){
  var bnames = ["B1","B2","B3","B4","B5","B7","AO","QA"];
  var bnumbers = [0,1,2,3,4,5,6,7];
```

```javascript
  var l5s = filterCollection(ee.ImageCollection('LEDAPS/LT5_L1T_SR'), startdate, enddate, poly)
          .select(bnumbers, bnames);
  return ee.ImageCollection(l5s);
}

//Stick atmospheric metadata to image as bands.
function addAtmosBands(srimg){
  // var ozone = ee.Image(srimg.get('ozone')).select([0], ['OZONE']);
  var tair = ee.Image(ee.List(srimg.get('surface_temp')).get(0))
          .select([0], ['SRTAIR00'])
          .addBands(ee.Image(ee.List(srimg.get('surface_temp')).get(1))
            .select([0], ['SRTAIR06']))
          .addBands(ee.Image(ee.List(srimg.get('surface_temp')).get(2))
            .select([0], ['SRTAIR12']))
          .addBands(ee.Image(ee.List(srimg.get('surface_temp')).get(3))
            .select([0], ['SRTAIR18']));
  var wv = ee.Image(ee.List(srimg.get('surface_wv')).get(0))
          .select([0], ['SRWVAP00'])
          .addBands(ee.Image(ee.List(srimg.get('surface_wv')).get(1))
            .select([0], ['SRWVAP06']))
          .addBands(ee.Image(ee.List(srimg.get('surface_wv')).get(2))
            .select([0], ['SRWVAP12']))
          .addBands(ee.Image(ee.List(srimg.get('surface_wv')).get(3))
            .select([0], ['SRWVAP18']));
  return srimg.addBands(tair).addBands(wv);
}

//Join Landsat collections based on system:time_start
function joinLandsatCollections(coll1, coll2){
  var eqfilter = ee.Filter.equals({'rightField':'system:time_start',
                        'leftField':'system:time_start'});
  var join = ee.Join.inner();
  var joined = ee.ImageCollection(join.apply(coll1, coll2, eqfilter));
  //Inner join returns a FeatureCollection with a primary and secondary set of
  //properties. Properties are collapsed into different bands of an image.
  return joined.map(function(element){
              return ee.Image.cat(element.get('primary'), element.get('secondary'));
            })
        .sort('system:time_start');
}

//Compute NDVI from a Landsat image.
function addNDVI(lndstimg){
  var ndvi = lndstimg.normalizedDifference(['B4', 'B3']);
  return lndstimg.addBands(ndvi.select([0],['NDVI']));
}

//-----------------------------------------------------------------------------------------------
//-----------------------------------------Thermal----------------------------------------------
//Compute emissivity from NDVI
//Note: NDVImin, NDVImax should be actually extracted from the image histogram.
//     Esoil, Eveg can be found in the literature.
function coreEmissivity(reflimg, NDVImin, NDVImax, Esoil, Eveg){
  var ndvi_min = ee.Image(ee.Number(NDVImin));
```

```
  var ndvi_max = ee.Image(ee.Number(NDVImax));
  var ndvi = reflimg.normalizedDifference(["B4", "B3"]);
  var fvc =  ndvi.subtract(ndvi_min)
        .divide(ndvi_max.subtract(ndvi_min))
        .pow(ee.Image(2));
  var e = ee.Image(Esoil).multiply(ee.Image(1).subtract(fvc)).add(ee.Image(Eveg).multiply(fvc));
  return e.select([0], ['emissivity']);
}

//Convenience function for mapping emissivity computation over collection
function getEmissivity(reflimg){
  return coreEmissivity(reflimg, 0.18, 0.85, 0.97, 0.99, 0.55);
}

//Compute psi functions
function getPsis(joinedimg){
  // WTR in NCEP data is in kg/m^2,
  // LST method needs g/cm^2: 1 kg/m2 = 10^-1 g/cm^2
  // NCEP values need to be unpacked first:
  //     + offset = 277.65
  //     + scale =0.01 for all images.
  var wv = joinedimg.select('SRWVAP18') //CAREFUL: TIME OF DAY HARDCODED
      .multiply(0.01) //scale
      .add(277.65) //offset
      .multiply(ee.Image(0.1)); //conversion to g/cm2
  var psi1 = ee.Image(0.14714).multiply(wv.pow(ee.Image(2)))
        .add(ee.Image(-0.15583).multiply(wv))
        .add(ee.Image(1.1234));
  var psi2 = ee.Image(-1.1836).multiply(wv.pow(ee.Image(2)))
        .add(ee.Image(-0.37607).multiply(wv))
        .add(ee.Image(-0.52894));
  var psi3 = ee.Image(-0.04554).multiply(wv.pow(ee.Image(2)))
        .add(ee.Image(1.8719).multiply(wv))
        .add(ee.Image(-0.39071));
  return ee.Image.cat([wv, psi1, psi2, psi3])
      .select([0,1,2,3], ["wv_gcm-2", "psi1", "psi2", "psi3"])
      .set({'system:time_start':joinedimg.get('system:time_start')});
}

//Compute surface temperature (output in degrees Kelvin)
function getSurfaceTemp(joinedimg){
  var brightemp = joinedimg.select('B6_BRT');
  var radtemp = joinedimg.select('B6_RAD');
  var c1 = ee.Image(1.19104); // W um^4 m^-2
  var c2 = ee.Image(14387.7); // um K
  var lambda = ee.Image(11.457); //um (effective wavelength of TM B6)
  var beta = ee.Image(1256); //K
  var gamma = radtemp.multiply(c2).divide(brightemp.pow(2))
          .multiply(radtemp.multiply(lambda.pow(4)).divide(c1)
              .add(lambda.pow(-1)))
          .pow(-1);
  var delta = brightemp.subtract(radtemp.multiply(gamma));
  var psis = getPsis(joinedimg);
  var e = getEmissivity(joinedimg);
```

```javascript
    var toctemp = gamma.multiply(psis.select('psi1').multiply(radtemp)
                    .add(psis.select('psi2'))
                    .divide(e)
                    .add(psis.select('psi3')))
            .add(delta);
    var sigma = 5.67e-8; //W/m2/K4
    var tocrad = ee.Image(sigma).multiply(e)
                .multiply(toctemp.pow(ee.Image(4)))
                .divide(ee.Image(Math.PI));
    return ee.Image.cat(toctemp, brightemp, e, tocrad, radtemp)
        .select([0,1,2,3,4], ['TOCtemp', 'TOAtemp', 'emiss', 'TOCrad', 'TOArad'])
        .set({
          'DATE_ACQUIRED':joinedimg.get('DATE_ACQUIRED'),
          'LANDSAT_SCENE_ID':joinedimg.get('LANDSAT_SCENE_ID'),
          'SUN_AZIMUTH':joinedimg.get("SUN_AZIMUTH"),
          'SUN_ELEVATION':joinedimg.get("SUN_ELEVATION"),
          'system:time_start':joinedimg.get('system:time_start'),
        });
}

//-----------------------------------------------------------------------------------------------------
//---------------------------------------------Application---------------------------------------------
//-------------Initialize base collections
var refpoly = ee.Geometry.Polygon(
      [[[24.771, -28.655],
        [24.771, -28.595],
        [24.87, -28.595],
        [24.87, -28.655]]]);
var year=1984; //repeat once more for 1993 and 2001
var start_date = ee.Date.fromYMD(year,6,1);
var end_date = ee.Date.fromYMD(year,6,30);     // check that there is indeed only one scene
                                                //during this interval
var jcoll = joinLandsatCollections(getLandsatTOA(start_date, end_date , refpoly),
                    getLandsatSR(start_date, end_date, refpoly)
                    .map(addAtmosBands));
jcoll = joinLandsatCollections(jcoll, getLandsatRAD(start_date, end_date , refpoly));

var tcoll = jcoll.map(function(jimg){return getSurfaceTemp(jimg)});

//---Test single image
var jimg = ee.Image(jcoll.first());

var timg_full = ee.Image(tcoll.first()).select('TOCtemp');
var timg = ee.Image(tcoll.first()).select('TOCtemp').clip(refpoly);

jimg = addNDVI(jimg);
//print(jimg);
var ndvi_full = jimg.select('NDVI');
var ndvi= jimg.select('NDVI').clip(refpoly);
//print(timg);

//---Visualization
Map.centerObject(refpoly);
var thpalette = ["000066", "00FFFF","FFFF00", "FF0000"];
var ndvipalette = ["ff0000", "00ff00"];
```

```
Map.addLayer(timg, {min:250, max:300, palette:thpalette}, "Surface Temperature (K)");
Map.addLayer(ndvi_full, {min:-1, max:1, palette:ndvipalette}, "NDVI");

// Export the image, specifying scale and region.
Export.image.toDrive({
  image: timg,
  description: 'temp'+year.toString(),
  scale: 30,
  region: refpoly
});

// Export the image, specifying scale and region.
//Export.image.toDrive({
  image: ndvi,
  description: 'ndvi'+year.toString(),
  scale: 30,
  region: refpoly
});

// Export the image, specifying scale and region.
Export.image.toDrive({
  image: timg_full,
  description: 'temp_full'+year.toString(),
  scale: 30
});

// Export the image, specifying scale and region.
Export.image.toDrive({
  image: ndvi_full,
  description: 'ndvi_full'+year.toString(),
  scale: 30
});
```

## Appendix 2

```r
# Data import
ndvi1984_full <- raster("ndvi_full1984.tif") # NDVI from entire Landsat scene
temp1984_full <- raster("temp_full1984.tif") # LST from entire Landsat scene
ndvi1984 <- raster("ndvi1984.tif") # NDVI from area of interest
temp1984 <- raster("temp1984.tif") # LST from area of interest

# Preprocessing
joint.1984 <- data.frame(values(ndvi1984_full), values(temp1984_full))
names(joint.1984) <- c("NDVI", "Temperature")
breaks <- seq(0.2, 1, 0.05)
joint.1984$bin <- .bincode(joint.1984$NDVI, breaks, TRUE, TRUE)

# Extraction of maximum and minimum LST
temp.max <- aggregate(joint.1984$Temperature, by = list(joint.1984$bin), max)
temp.min <- aggregate(joint.1984$Temperature, by = list(joint.1984$bin), min)

# Linear regression of the scatter plot of max/min temperatures
bound.upper<-lm(temp.max$x ~ breaks[temp.max$Group.1])
bound.lower<-lm(temp.min$x ~ breaks[temp.min$Group.1])

# Plotting Linear regression results
png('joint1984.png')
plot(joint.1984$NDVI, joint.1984$Temperature, xlab="NDVI June 1984", ylab=expression("Temperature ["*~degree*K*"]"))+ abline(bound.upper,col="red")+ abline(bound.lower, col="blue")

# Estimation of max/min temperature per pixel
tsmax <- bound.upper$coefficients[2]*ndvi1984+bound.upper$coefficients[1]
tsmin <- bound.lower$coefficients[2]*ndvi1984+bound.lower$coefficients[1]

# Estimation of SMI per pixel
smi_1984 <- (tsmax-temp1984)/(tsmax-tsmin)

# Plotting SMI
plot(smi_1984)
```